\documentclass[11pt]{article}

\title{Credit vs. Discount-Based Congestion Pricing: A Comparison Study}
\author{Chih-Yuan Chiu\footnote{University of California, Berkeley, Department of Electrical Engineering and Computer Sciences}, Devansh Jalota\footnote{Stanford University, Institute for Computational \& Mathematical Engineering}, Marco Pavone\footnote{Stanford University, Department of Aeronautics and Astronautics}}

\usepackage{amsmath, amssymb}
\usepackage[utf8]{inputenc}
\usepackage{amsthm}

\allowdisplaybreaks

\usepackage[usenames, dvipsnames]{xcolor}
\usepackage{graphicx}
\usepackage{bbm}
\usepackage[inline]{enumitem}
\usepackage{url}
\usepackage[colorlinks=true,
raiselinks=true,
linkcolor=Sepia,
citecolor=MidnightBlue,
urlcolor=Sepia]{hyperref}

\usepackage[ruled,vlined,linesnumbered]{algorithm2e}
\usepackage[noend]{algpseudocode}

\usepackage{geometry}
\usepackage{siunitx}

\usepackage{mathtools}
\mathtoolsset{mathic=true}
\usepackage{todonotes}
\usepackage[varqu, varl, scale=0.9]{inconsolata}
\usepackage{booktabs}
\usepackage[normalem]{ulem}
\usepackage{dirtytalk}

\numberwithin{equation}{section}

\newtheorem{theorem}{Theorem}[section]

\newtheorem{proposition}[theorem]{Proposition}

\newtheorem{remark}{Remark}
\newtheorem{assumption}{Assumption}

\newtheorem{definition}[theorem]{Definition}

\DeclarePairedDelimiterX\loro[1]\rbrack\lbrack{#1}
\DeclarePairedDelimiterX\lorc[1]\rbrack\rbrack{#1}
\DeclarePairedDelimiterX\lcro[1]\lbrack\lbrack{#1}
\DeclarePairedDelimiterX\lcrc[1]\lbrack\rbrack{#1}
\DeclarePairedDelimiterX\set[1]\lbrace\rbrace{#1}

\newcommand{\ra}{\ensuremath{\rightarrow}}
\newcommand{\Ra}{\ensuremath{\;\Rightarrow\;}}

\DeclarePairedDelimiterX\norm[1]\lVert\rVert{#1}

\newcommand{\R}{\ensuremath{\mathbb{R}}}
\newcommand{\N}{\ensuremath{\mathbb{N}}}

\newcommand{\G}{\mathcal{G}}
\newcommand{\network}{\mathcal{N}}
\newcommand{\Y}{\mathcal{Y}}
\newcommand{\routes}{\textbf{R}}
\newcommand{\nodes}{I}
\newcommand{\edges}{E}

\newcommand{\revision}[1]{{\textcolor[rgb]{0,0,0}{#1}}}

\usepackage{todonotes}

\begin{document}

\maketitle

\begin{abstract}
Congestion pricing offers a promising traffic management policy for regulating congestion, but has also been criticized for placing outsized financial burdens on low-income users. Credit-based congestion pricing (CBCP) and discount-based congestion pricing (DBCP) policies, which respectively provide travel credits and toll discounts to subsidize low-income users' access to tolled roads, are promising mechanisms for reducing traffic congestion without worsening societal inequities. However, the optimal design and relative merits of CBCP and DBCP policies remain poorly understood. This work studies the effects of deploying CBCP and DBCP policies to route users on multi-lane highway networks with tolled express lanes. We formulate a non-atomic routing game in which a subset of \emph{eligible} users is granted toll relief via a fixed budget or toll discount, while the remaining \emph{ineligible} users must pay out-of-pocket. We prove that Nash equilibrium traffic flow patterns exist under any CBCP or DBCP policy. Under the additional assumption that eligible users have time-invariant values of time (VoTs), we provide a convex program to efficiently compute these equilibria. Moreover, for single-edge networks, we identify conditions under which DBCP policies outperform CBCP policies, in the sense of improving eligible users' express lane access\revision{, an equity objective often neglected in existing congestion pricing schemes that price low-income users out of utilizing express lanes. Specifically, we identify user and network-dependent parameters whose values play a key role in determining whether DBCP or CBCP policies are more effective at expanding eligible users' express lane access.} Finally, we present empirical results from a CBCP pilot study of the San Mateo 101 Express Lane Project in California. Our empirical results corroborate our theoretical analysis of the impact of deploying credit-based and discount-based policies, and lend insights into the sensitivity of their impact with respect to the travel demand and users' VoTs.
\end{abstract}

\section{Introduction}
\label{sec: Introduction and Related Works}

In recent years, increasing levels of traffic congestion in urban areas have spurred the development of effective policies aimed at regulating traffic flow, shortening commute times, and reducing air pollution.
Congestion pricing, which applies monetary costs to roads to reshape user incentives and promote more efficient use of traffic infrastructure, offers a promising traffic control mechanism. Both fundamental research \cite{Cole2003HeterogeneousTolling, MaheshwariKulkarni2022DynamicTollingforInducingSociallyOptimalTrafficLoads,  Florian2003NetworkEquilibriumAndPricing} and real-world empirical studies established that congestion pricing can alleviate congestion in societal-scale transportation systems \cite{Manville2018WouldCongestionPricingHarmThePoor, Ley2023NYCCongestion, Maheshwari2024CongestionPricingforEfficiencyEquity}. Although real-life implementations of congestion pricing over entire traffic networks remain rare, many urban centers impose such policies on highway express lanes, to provide travelers with more efficient travel options during periods of high congestion.

Unfortunately, despite their promise, existing congestion pricing schemes have also raised serious equity concerns. In particular, current designs for express lanes have been criticized for serving as \say{elitist Lexus lanes}, offering shorter commute times to wealthier users at the cost of extending the commute times of low-income users \cite{Klein2021AreTollLanesElististOrProgressive}. To address the inequity issues raised by existing policies, both \textit{credit-based} congestion pricing (CBCP) and \textit{discount-based} congestion pricing (DBCP) policies have been proposed as mechanisms to alleviate congestion without exacerbating existing societal inequities. Credit-based policies offer lower-income commuters travel assistance via a fixed budget, while discount-based policies provide toll discounts \cite{Jalota2023CreditBasedCongestionPricing, Kockelman2005CBCP}. For example, in California, the San Mateo 101 Express Lanes Equity Program recently initiated a CBCP policy, to ease the financial strain that congestion fees place on low-income communities \cite{CBCP-SanMateo}. Elsewhere in the San Francisco area, the Express Lanes START$^{\text{SM}}$ program recently announced a DBCP policy, offering low-income drivers access to Interstate 880 express lanes at discounted toll rates \cite{DBCP-ExpressLaneSTART}. 

However, despite the growing attention that both CBCP and DBCP mechanisms have attracted recently, limited research has been conducted to understand their strengths and drawbacks. To address this gap, our work contrasts the effects of deploying CBCP and DBCP policies to route users with heterogeneous values of time (VoT), on a highway network with tolled express lanes. \revision{Specifically, we examine the effects of deploying CBCP and DBCP policies on low-income users' express lane access. We focus on the low-income users' express lane access as our \emph{equity metric}, as conventional congestion pricing schemes typically price low income users out of express lanes. We demonstrate that the relative success of CBCP and DBCP policies in improving low-income users' express lane access hinges upon a number of user-specific and network-specific properties, such as user VoTs and latency functions that map flow levels to travel times on network edges. Thus, our work provides guidance to policy makers, by pinpointing user and network attributes which must be accurately determined in order to design effective traffic regulations.
}

\revision{The remainder of this paper is structured as follows. Section \ref{sec: Related Work} surveys relevant literature on credit-based and discount-based congestion pricing}. We present a mixed-economy model, in which eligible users receive travel assistance in the form of credits or toll discounts, while ineligible users pay entirely out-of-pocket to access express lanes. Then, in Section \ref{sec: DBCP and CBCP Equilibria}, we formally introduce the operation of DBCP and CBCP policies and associated user costs, as well as Nash equilibrium concepts for the corresponding steady-state traffic flow patterns, which we call CBCP and DBCP equilibria, respectively. Our main contributions include:
\begin{enumerate}
    \item In Section \ref{sec: Equilibria Characterization}, we prove that CBCP and DBCP equilibria exist. We also present convex optimization problems for efficiently computing these equilibria, under the assumption that eligible users' VoTs are time-invariant. 
    
    \item In Section \ref{sec: Budget vs Discount}, for the setting where the network consists of a single multi-lane highway, we contrast the effects of CBCP and DBCP policies that represent equivalent levels of allocated travel assistance (Remark \ref{Remark: CBCP and DBCP, Fair Comparison}).

    \item In Section \ref{sec: Experimental Results}, we perform sensitivity analysis on the theoretical results in Section \ref{sec: Budget vs Discount}, and study the effects of deploying CBCP and DBCP policies in the context of the San Mateo County 101 Express Lanes Project.
\end{enumerate}
Finally, in Section \ref{sec: Conclusion and Future Work}, we summarize our findings and discuss directions for future research. 

\paragraph{Notation}
For each positive integer $N \in \N$, we define $[N] := \{1, \cdots, N\}$. We define $\textbf{1}\{\cdot\}$ to be the indicator function that returns 1 when the input argument is true, and 0 otherwise.

\section{Related Work}
\label{sec: Related Work}

Congestion pricing has been extensively studied as a promising mechanism to alleviate traffic congestion \cite{Pigou1912WealthAndWelfare, Roughgarden2010AlgorithmicGameTheory}. Many recent works have focused on designing tolling schemes that address equity as well as efficiency concerns \cite{Jalota2022Balancing, Jalota2021WhenEfficiencyMeetsEquity, Maheshwari2024CongestionPricingforEfficiencyEquity}. While the literature on equitable congestion pricing mechanisms is vast, below we review related work on two categories in particular: credit-based congestion pricing and discount-based congestion pricing.

Credit-based congestion pricing (CBCP) provides a travel budget to low-income users, to reduce the financial burden of tolls
\cite{Li2020TrafficAndWelfareImpactsOfCBCP, Jalota2023CreditBasedCongestionPricing}. The study of CBCP policies for traffic control has largely focused on assigning \textit{tradeable} credits, which can be used by beneficiaries to access alternative transport services, such as public transit. Recently, Jalota et al. \cite{Jalota2023CreditBasedCongestionPricing} studied a \textit{non-tradeable} CBCP scheme for controlling traffic on a single multi-lane highway with an express lane, within a mixed economy framework. In particular, the CBCP scheme in \cite{Jalota2023CreditBasedCongestionPricing} provides eligible users with credits that can only be used to access express lanes, and cannot be exchanged for any other good or service. Our work shares with \cite{Jalota2023CreditBasedCongestionPricing} a focus on designing non-tradeable CBCPs. However, we introduce a novel notion of CBCP equilibria for the setting in which eligible users on the express lane can pay tolls out of pocket. Moreover, unlike \cite{Jalota2023CreditBasedCongestionPricing}, we also characterize equilibrium flows under DBCP policies, and study the effects of deploying CBCP and DBCP policies on general networks. Since the travel credits considered in our framework are non-tradeable, our work is also linked to the literature on artificial currency schemes \cite{Budish2011CombinatorialAssignmentProblem}. However, unlike most artificial currency mechanisms, our method considers a mixed economy setting in which only a subset of users are allotted credits.

In contrast to CBCPs, discount-based congestion pricing (DBCP) instead offers disadvantaged users toll discounts \cite{DBCP-ExpressLaneSTART, London2024ResidentsDiscount}. DBCPs impose differentiated tolls on different user groups, and are thus closely aligned to the well-established literature on heterogeneous tolling schemes \cite{Fleischer2004TollsForHeterogeneousSelfishUsers, Guo2010ParetoImprovingCongestionPricing}, which differentiate users based on various attributes, such as whether the user's vehicle is self-driving \cite{Mehr2019PricingTrafficNetworksWithMixedVehicleAutonomy, Lazar2019OptimalTollingforHeterogeneousTrafficNetworkswithMixedAutonomy} or uses clean energy \cite{Perera2020MulticlassTollBasedApproach}. 
However, our method specifically subsidizes low-income users, to address equity concerns in existing congestion pricing mechanisms. Moreover, we contrast the effects of deploying DBCP and CBCP policies on express lane usage at equilibrium. 

In terms of methodology, the equilibrium flow analysis in our work closely aligns with the rich literature on Nash equilibrium concepts for routing games \cite{Wardrop1952SomeTheoreticalAspectsOfRoadTrafficResearch, Roughgarden2010AlgorithmicGameTheory}, particularly \cite{Jalota2023CreditBasedCongestionPricing}, which first described equilibrium flows under CBCP policies deployed on single-edge networks with a tolled express lane and an untolled general express lane.

\section{DBCP and CBCP Equilibria}
\label{sec: DBCP and CBCP Equilibria}

Here, we present the network and traffic flow model (Sec. \ref{subsec: Setup}), the DBCP and CBCP policies, 
and the notions of equilibrium flow (Sec. \ref{subsec: DBCP}, \ref{subsec: CBCP}) studied in this work.

\subsection{Setup}
\label{subsec: Setup}

We study the design of DBCP and CBCP policies on a transportation network, where each edge of the network consists of an express lane that can be tolled and a general purpose lane that cannot be tolled.
Formally, let $\network = (\nodes, \edges)$ be an (acylic) traffic network, where $\nodes$ and $\edges$ are the set of nodes and the set of edges in $\network$, respectively. For each node $i \in \nodes$, let $E_i^-$ and $E_i^+$ respectively denote the set of incoming edges and outgoing edges at node $i$. For ease of exposition, we assume $\network$ contains only one origin-destination pair ($o, d$), though our methods extend to multi-origin multi-destination networks. Each edge $e \in \edges$ contains an \textit{express} toll lane $(k = 1)$ and a \textit{general purpose} lane $(k = 2)$. Each lane $k \in [2]$ on each edge $e \in \edges$ is associated with a differentiable, strictly positive, strictly increasing, strictly convex latency function $\ell_{e,k}$, which maps the flow level to the travel time on edge $e$, lane $k$. 

At each time $t$ in a finite time horizon $[T] := \{1, \cdots, T\}$, each user travels through the network $\network$ from origin to destination. A subset of users receives subsidies for express lane use, via a budget or a toll discount. As is standard in the literature, we partition users into a finite set $\G$ of user groups, based on their income level, values of time (VoTs), and eligibility for travel subsidies. Let $\G_E$ and $\G_I$ be the sets of \textit{eligible} and \textit{ineligible} user groups, respectively. Let $v_{t,g}$ denote the VoT of users in the group $g \in \G$ at time $t \in [T]$, i.e., the monetary value each group $g$ user is willing to pay for a unit of reduced travel time. For ease of exposition, we assume the total travel demand (i.e., the users' total flow) of each group $g$ in the network is fixed and normalized to one.

\subsection{Discount-Based Congestion Pricing (DBCP)}
\label{subsec: DBCP}

A DBCP policy is 
described by
a tuple ($\boldsymbol{\tau}$, $\boldsymbol{\alpha}$), where $\boldsymbol{\tau} := (\tau_{e,t})_{e \in \edges, t \in [T]} \in \R_{\geq 0}^{|\edges|T}$ are tolls imposed at each time on the express lane along each edge in the network, while $\boldsymbol{\alpha} := (\alpha_{e,t})_{e \in \edges, t \in [T]} \in [0, 1]^{|\edges|T}$ are toll discounts at each time along each edge. Under the DBCP policy, each ineligible user on the express lane ($k=1$) on any edge $e \in \edges$ must pay the full toll $\tau_{e,t}$, while each eligible user taking the same lane must pay the \textit{discounted} toll $(1 - \alpha_{e,t}) \tau_{e,t}$.

Below, given a $(\boldsymbol{\tau}, \boldsymbol{\alpha})$-DBCP policy, we first define an admissible set of flow patterns, which specifies flow continuity and non-negativity constraints (Sec. \ref{subsubsec: Flow Constraints, DBCP}). We then describe the travel and toll costs associated with the express and general purpose lane on each edge in the network for both eligible and ineligible users (Sec. \ref{subsubsec: Travel and Toll Costs, DBCP}). Finally, we define the corresponding $(\boldsymbol{\tau}, \boldsymbol{\alpha})$-DBCP equilibrium (Sec. \ref{subsubsec: DBCP Equilibria}).

\subsubsection{Flow Constraints under DBCP Policies}
\label{subsubsec: Flow Constraints, DBCP}

To characterize the equilibrium lane flows that result from users' selfish route choices under a DBCP policy, let $y_t^g := (y_{e,k,t}^g)_{e \in \edges, k \in [2]} \in \R_{\geq 0}^{2|\edges|}$ denote the flow of users in each group that is routed onto each edge and lane at each time. We define the set of feasible flows as:
{
\begin{alignat}{2} \label{Eqn: Ytgd, Continuity of flow}
    \Y_t^{g,d} := &\Bigg\{ y_t^g \in \R^{2|E|}: \hspace{5mm} &&\sum_{\hat e \in \edges_i^+} \sum_{k = 1}^2 y_{e,k,t}^g = \textbf{1}\{i=o \} + \sum_{\hat e \in \edges_i^-} \sum_{k = 1}^2 y_{e,k,t}^g,  \ \forall \ i \in \nodes \backslash \{d\}, \\ \label{Eqn: Ytgd, Non-negativity of flow}
    & &&y_{e,k,t}^g \geq 0, \ \forall \ e \in \edges, k \in [2]. \Bigg\}
\end{alignat}
}
Above, \eqref{Eqn: Ytgd, Continuity of flow} are allocation constraints that encode flow continuity at each node $i \in \nodes \backslash \{o, d\}$, and ensure that the total traffic flow routed from the origin node $o$ to the destination node equals the (normalized) demand. Meanwhile, \eqref{Eqn: Ytgd, Non-negativity of flow} constrains all flows to be non-negative.

The set of feasible flows corresponding to all users, eligible and ineligible, is given by:
\begin{align} \label{Eqn: Y, Discount}
    \Y^d := \prod_{t=1}^T \prod_{g \in \G} \Y_t^{g,d}.
\end{align}
Below, let $\textbf{y} := (y_{e,k,t}^g)_{g \in \G, e \in \edges, k \in [2], t \in [T]} \in \R^{2|E|T}$ denote the flow pattern for each user group over the time horizon $T$, and let $\textbf{x}$ denote the aggregate lane flows  corresponding to $\textbf{y}$ by $x_{e,k,t} := \sum_{g \in \G} y_{e,k,t}^g$, for each group $g \in \G$.

\subsubsection{Travel and Toll Costs under DBCP Policies}
\label{subsubsec: Travel and Toll Costs, DBCP}

The travel cost of each user is given by weighted sums of their travel times and tolls. Under the discount-based policy, we define the cost per user from group $g \in \G$ as follows: At each time $t \in [T]$, for each edge $e \in \edges$ and lane $k \in [2]$:
\begin{align*}
    &c_{e,k,t}^g(\textbf{y}) := \ \begin{cases}
        v_t^g \ell_{e,1}(x_{e,1,t}) + \tau_{e,t}, \hspace{3.5cm} &k = 1, g \in \G_I, \\
        v_t^g \ell_{e,1}(x_{e,1,t}) + (1 - \alpha_{e,t}) \tau_{e,t}, &k = 1, g \in \G_E, \\
        v_t^g \ell_{e,2}(x_{e,2,t}), &k = 2.
    \end{cases}
\end{align*}
In words, at each time $t$, on each edge $e \in \edges$, each ineligible traveler from group $g \in \G_I$ who accesses the express lane incurs the full toll value $\tau_{e,t}$ and the travel cost $v_t^g \ell_{e,1}(x_{e,1,t})$, while each eligible traveler from group $g \in \G_E$ incurs the reduced toll value $(1 - \alpha_{e,t}) \tau_{e,t}$ and the travel cost $v_t^g \ell_{e,1}(x_{e,1,t})$. Each traveler  on the general purpose lane, eligible or ineligible, incurs the same cost $v_t^g \ell_{e,2}(x_{e,2,t})$.

\subsubsection{DBCP Equilibria}
\label{subsubsec: DBCP Equilibria}

At each time $t \in [T]$, every individual user selfishly selects a route (i.e., edge-lane sequence) connecting the origin $o$ to the destination $d$ that minimizes the overall cost, among all feasible routes. These route selection decisions, in aggregate, form the flows $\textbf{y}$ on each edge and lane of the network. A Nash equilibrium flow $\textbf{y}^\star \in \Y^d$ is then defined as a flow pattern in which only cost-minimizing routes have strictly positive flows. However, below, we use an equivalent definition formulated using variational inequalities involving the flows on each edge and lane \cite{Roughgarden2010AlgorithmicGameTheory}, without referencing routes. In particular, for a given DBCP policy $(\boldsymbol{\tau}, \boldsymbol{\alpha})$, a flow pattern $\textbf{y}^\star$ is called a DBCP equilibrium if no user can reduce their travel cost by unilaterally deviating from their route selections.

\begin{definition}[\textbf{DBCP Equilibrium}] \label{Def: DBCP Equilibria}
We call $\textbf{y}^\star \in \Y^d$ a ($\boldsymbol{\tau}$, $\boldsymbol{\alpha}$)-DBCP equilibrium corresponding to the ($\boldsymbol{\tau}$, $\boldsymbol{\alpha}$)-DBCP policy if, for group $g \in \G, t \in [T]$, and any $y_t^g \in \Y_t^{g,d}$, we have $\sum_{e \in \edges} \sum_{k=1}^2 (y_{e,k,t}^g - y_{e,k,t}^{g \star}) c_{e,k,t}^g(\textbf{y}^\star) \geq 0$.
\end{definition}

Note that if $\alpha_{e,t} = 0, \forall e \in \edges$, $t \in [T]$, then all users are ineligible, and the DBCP equilibrium is the Nash equilibrium for non-atomic travelers with heterogeneous VoTs.

\subsection{Credit-Based Congestion Pricing (CBCP)}
\label{subsec: CBCP}

A CBCP policy is 
described by 
a tuple ($\boldsymbol{\tau}$, $B$), where $\boldsymbol{\tau} := \{(\tau_{e,t})_{e \in \edges, t \in [T]} \} \in \R_{\geq 0}^{|\edges|T}$ is the vector of tolls imposed on express lanes at each edge throughout the network, while $B$ denotes the total travel credit (i.e., budget) given to each eligible user.

Below, we first define an admissible set of flow patterns corresponding to a given $(\boldsymbol{\tau}, B)$-CBCP scheme, which specifies flow continuity, non-negativity, and budget constraints (Sec. \ref{subsubsec: Flow Constraints, CBCP}). We then describe the travel and toll costs associated with the express and general purpose lane on each edge in the network for both eligible and ineligible users (Sec. \ref{subsubsec: Travel and Toll Costs, CBCP}). Finally, we define the corresponding $(\boldsymbol{\tau}, B)$-CBCP equilibrium (Sec. \ref{subsubsec: CBCP Equilibria}).

\subsubsection{Flow Constraints Under CBCP Policies}
\label{subsubsec: Flow Constraints, CBCP}

As is the case under DBCP policies, ineligible users can only access express lanes by paying the entire toll out of pocket. However, at each time $t$, each eligible user taking the express lane ($k=1$) on any edge $e \in \edges$ can pay the toll $\tau_{e,t}$, using a combination of their available budget and out-of-pocket funds. Accordingly, at each time $t$, on any edge $e \in \edges$, we use $\tilde y_{e,1,t}^g$ (resp., $\hat y_{e,1,t}^g$) to model the flow level of users from group $g \in \G$ who use part of their budget (resp., pays out of pocket) to access the express lane on edge $e$. Since ineligible users have zero budget, we have $\tilde y_{e,1,t}^g = 0$ for each $e \in \edges$, $t \in [T]$, $g \in \G_I$. As before, let $y_{e,2,t}^g$ denote the flow level of users from group $g$ on the general purpose lane on edge $e$ at time $t$. Finally, let $y_t^g := (\tilde y_{e,1,t}^g, \hat y_{e,1,t}^g, y_{e,2,t}^g)_{e \in \edges} \in \R_{\geq 0}^{3|\edges|}$ denote the flow pattern of users in group $g \in \G$ routed onto each edge at time $t \in [T]$.

\begin{remark}
Our model differs from the CBCP scheme in Jalota et al. \cite{Jalota2023CreditBasedCongestionPricing}, which only considers single-edge networks with a tolled express lane and an untolled general purpose lane, and requires eligible users to pay tolls only using their budget. In Jalota et al. \cite{Jalota2023CreditBasedCongestionPricing}, eligible users only decide the fraction of flow to be routed on the express lane.
In our model, eligible users on express lanes must also decide what fraction of the toll to pay using their budget. To account for this additional level of decision-making, we introduce separate terms to describe the fraction of eligible users' flow  whose corresponding toll costs are covered by their available budget ($\tilde y_{e,1,t}^g$), and the remaining fraction whose toll costs are paid using personal funds ($\hat y_{e,1,t}^g$).
\end{remark}

Under CBCP policies, eligible groups receive toll subsidies via a fixed budget over the time horizon $T$, while ineligible groups do not. Thus, constraint sets for eligible and ineligible groups' flows are defined separately. For each eligible group $g \in \G_E$, we define the set $\Y^{g,b}$ of feasible flows 
$y_t^g \in \R^{3|E|}$ 
for group $g$ over the time horizon $T$ as:
{
\begin{align} 
\nonumber
    \Y^{g,b} := &\Big\{ y^g \in \R^{3|\edges|T}:  
    \tilde y_{e,1,t}^g + \hat y_{e,1,t}^g = y_{e,1,t}^g, \forall e \in \edges, t \in [T], 
    \\ \label{Eqn: Ygb, Continuity of flow}
    &\hspace{5mm} \sum_{\hat e \in \edges_i^+} \sum_{k = 1}^2 y_{e,k,t}^g = \textbf{1}\{i=o \} + \sum_{\hat e \in \edges_i^-} \sum_{k = 1}^2 y_{e,k,t}^g, \ \forall \ i \in \nodes \backslash \{d\}, t \in [T], \\ \label{Eqn: Ygb, Non-negativity of flow}
    &\hspace{5mm} \tilde y_{e,1,t}^g, \hat y_{e,1,t}^g, y_{e,2,t}^g \geq 0, \ \forall \ e \in \edges, t \in [T], \\ \label{Eqn: Ygb, Budget Constraint}
    &\hspace{5mm} \sum_{t \in [T]} \sum_{e \in \edges} \tilde y_{e,1,t}^g \tau_{e,t} \leq B. \Big\}.
\end{align}
}
Above, 
\eqref{Eqn: Ygb, Continuity of flow} encodes flow continuity at each node, while \eqref{Eqn: Ygb, Non-negativity of flow} constrains all flows to be non-negative. Finally, \eqref{Eqn: Ygb, Budget Constraint} enforces the budget constraint.

\begin{remark}
The budget constraint \eqref{Eqn: Ygb, Budget Constraint} generalizes the analogous notion for CBCP equilibrium flows on single-edge, two-link networks, given in Jalota et al. \cite{Jalota2023CreditBasedCongestionPricing}, to the general network setting. When a CBCP policy is deployed on a general network, the budget constraint applies to each route in the network. Specifically, let $\routes$ denotes the set of all routes, where each route is a finite sequence of edge-lane tuples $\{ (e_1, k_1), \cdots, (e_m, k_m) \}$ connecting the origin and destination, where $k_i \in \{\tilde{1}, \hat{1}, 2\}$, with $\tilde{1}$ and $\hat{1}$ indicating express lane toll payment via the given budget and via out-of-pocket funds, respectively. Let $y_{r,t}^g$ denote the flow of an eligible group $g$ on route $r$ at time $t$. Then the budget constraint is:
\begin{align} \label{Eqn: Budget constraint, Natural}
    \sum_{t \in [T]} \sum_{r \in \routes} y_{r,t}^g \sum_{e \in \edges: (e, \tilde{1}) \in r} \tau_{e,t} \leq B.
\end{align}
However, \eqref{Eqn: Budget constraint, Natural} can still be formulated solely in terms of the edges $\edges$ of the network, without reference to routes:
{
\begin{align*}
    \sum_{t \in [T]} \sum_{r \in \routes} y_{r,t}^g \sum_{e \in \edges: (e, \tilde{1}) \in r} \tau_{e,t} = \hspace{0.5mm} &\sum_{t \in [T]} \sum_{e \in \edges} \tau_{e,t}\sum_{r \in \routes: (e, \tilde{1}) \in r} y_{r,t}^g = \sum_{t \in [T]} \sum_{e \in \edges} \tau_{e,t} \tilde y_{e,1,t}^g.
\end{align*}
}
The first equality above follows by reordering the summations over routes and over edges, while the second follows by noting an edge flow is the sum of all route flows passing through that edge \cite{Roughgarden2010AlgorithmicGameTheory}. Thus, \eqref{Eqn: Budget constraint, Natural} can be written as \eqref{Eqn: Ygb, Budget Constraint}.
\end{remark}

Since ineligible user groups are not given any travel credit or toll discount, 
their flows are not restricted by the budget constraints \eqref{Eqn: Ygb, Budget Constraint}. As a result, the constraint set for ineligible groups' flows decouples across time. Specifically, at each time $t$, for each ineligible group $g \in G_I$, we define the set $\Y_t^{g,b}$ of feasible flow patterns for group $g$ as:
{
\begin{alignat}{2} \nonumber
    \Y_t^{g,b} := &\Bigg\{ y_t^g \in \R^{3|E|}: \hspace{5mm} &&\tilde y_{e,1,t}^g = 0, \forall \ e \in \edges, \\ \nonumber
    & &&\sum_{e' \in \edges_i^+} \sum_{k = 1}^2 y_{e',k,t}^g = \textbf{1}\{i=o \} + \sum_{\hat e \in \edges_i^-} \sum_{k = 1}^2 y_{\hat e,k,t}^g, \ \forall \ i \in \nodes \backslash \{d\}, \\ \nonumber
    & &&\tilde y_{e,1,t}^g, \hat y_{e,1,t}^g, y_{e,2,t}^g \geq 0, \ \forall \ e \in \edges. \Bigg\}
\end{alignat}
}
Let $\textbf{y} := (y^g \in \Y^{g,b}: g \in G_E, y_t^g \in \Y_t^{g,b}: g \in G_I, t \in [T])$. The feasible flow set of all eligible and ineligible user groups is:
{
\begin{align} \label{Eqn: Y, Credit}
    \Y^b := \Bigg( \prod_{g \in \G_E} \Y^{g,b} \Bigg) \times \Bigg( \prod_{t=1}^T \prod_{g \in \G_I} \Y_t^{g,b} \Bigg),
\end{align}
}
Again, we define the lane flows $\textbf{x}$ corresponding to $\textbf{y}$ by $x_{e,k,t} := \sum_{g \in \G} y_{e,k,t}^g$, for each group $g \in \G$.

\subsubsection{Travel and Toll Costs under CBCP Policies}
\label{subsubsec: Travel and Toll Costs, CBCP}

Under the $(\boldsymbol{\tau}, B)$-CBCP policy, the cost incurred per user from group $g \in \G$ is as defined below. For each group $g \in \G$, on each edge $e \in \edges$, at each time $t \in [T]$:
\begin{align*}
    \tilde c_{e,1,t}^g(\textbf{y}) &:= v_t^g \ell_{e,1}(x_{e,1,t}), \\
    \hat c_{e,1,t}^g(\textbf{y}) &:= v_t^g \ell_{e,1}(x_{e,1,t}) + \tau_{e,t} \\
    c_{e,2,t}^g(\textbf{y}) &:= v_t^g \ell_{e,2}(x_{e,2,t}).
\end{align*}
In words, at each time $t$, on each edge $e \in \edges$, each ineligible traveler from group $g \in \G_I$ who uses the express lane incurs the full toll cost $\tau_{e,t}$ and the travel cost $v_t^g \ell_{e,1}(x_{e,1,t})$ (recall that $\tilde y_{e,1,t}^g = 0$ for each $g \in G_I$, since ineligible users have zero budget). Meanwhile, each eligible traveler from group $g \in \G_E$ who uses the express lane can pay the toll using a mix of their available budget and personal funds, thus incurring a convex combination of the travel cost $\tilde c_{e,1,t}^g(\textbf{y}) := v_t^g \ell_{e,1}(x_{e,1,t})$ and the full travel-plus-toll cost $\hat c_{e,1,t}^g(\textbf{y}) := v_t^g \ell_{e,1}(x_{e,1,t}) + \tau_{e,t}$. Each traveler incurs the same cost $v_t^g \ell_{e,2}(x_{e,2,t})$ when accessing the general purpose lane, regardless of whether they are eligible or ineligible.

\subsubsection{CBCP Equilibria}
\label{subsubsec: CBCP Equilibria}

We formulate the Nash equilibrium flows under CBCP policies as follows. 

\begin{definition}[\textbf{CBCP Equilibrium}] \label{Def: CBCP Equilibria}
We call $\textbf{y}^\star \in \Y^b$ a ($\boldsymbol{\tau}$, $B$)-CBCP equilibrium corresponding to the ($\boldsymbol{\tau}$, $B$)-CBCP policy if, for any $\textbf{y} \in \Y^b$, for each ineligible group $g \in \G_I$ and time $t \in [T]$:
{
\begin{align*}
    &\sum_{e \in \edges} \sum_{k=1}^2 \sum_{t=1}^T (\hat y_{e,k,t}^g - \hat y_{e,k,t}^{g \star}) \hat c_{e,k,t}^g (\textbf{y}^\star) \geq 0,
\end{align*}
}
and for each eligible group $g \in \G_E$:
{
\begin{align*}
    &\sum_{t \in [T]} \sum_{e \in \edges} \big( (\tilde y_{e,1,t}^g - \tilde y_{e,1,t}^{g \star}) \tilde c_{e,1,t}^g (\textbf{y}^\star) + (\hat y_{e,1,t}^g - \hat y_{e,1,t}^{g \star}) \hat c_{e,1,t}^g (\textbf{y}^\star) + (y_{e,2,t}^g - y_{e,2,t}^{g \star}) \hat c_{e,2,t}^g (\textbf{y}^\star) \big) \geq 0.
\end{align*}
}
\end{definition}

Note that if $B = 0$, then all users are ineligible, and the CBCP equilibrium reduces to the Nash equilibrium for non-atomic travelers with heterogeneous VoTs.

\section{Equilibria Characterization}
\label{sec: Equilibria Characterization}

Below, we analyze properties of DBCP and CBCP equilibria to study the impact of deploying DBCP and CBCP policies on traffic systems. In Sec. \ref{subsec: Existence of DBCP and CBCP Equilibria}, we prove that DBCP (respectively, CBCP) equilibria exist given any DBCP policy (respectively, CBCP) policy. Then, in Sec. \ref{subsec: Convex Program Characterization}, we show that if eligible users' VoTs are fixed in time, 
DBCP and CBCP equilibria
can be efficiently computed by solving a convex program. Finally, in Sec. \ref{subsec: Uniqueness of Edge Flows}, we prove that given a DBCP (respectively, CBCP) policy, although the corresponding DBCP (respectively, CBCP) equilibria need not be unique, the total lane flow at equilibrium is unique.


\subsection{Existence of DBCP and CBCP Equilibria}
\label{subsec: Existence of DBCP and CBCP Equilibria}

Below, we use the variational inequalities in the definition of DBCP equilibra (Definition \ref{Def: DBCP Equilibria}) and CBCP equilibria (Definition \ref{Def: CBCP Equilibria}) to establish the existence of these equilibria.

\begin{proposition} \label{Prop: Lane Flows Existence, DBCP}
For each $(\boldsymbol{\tau}, \boldsymbol{\alpha})$-DBCP policy, where $\boldsymbol{\tau}$ is component-wise non-negative and $\boldsymbol{\alpha}$ is component-wise in $[0, 1]$, there exists a $(\boldsymbol{\tau}, \boldsymbol{\alpha})$-DBCP  equilibrium.
\end{proposition}

\begin{proof}
The flow constraint set $\Y$ is compact and the cost functions $c_{e,k,t}^g$ are continuous. Thus, the existence of $(\boldsymbol{\tau}, \boldsymbol{\alpha})$-DBCP  equilibria follow from the theory of variational inequalities \cite{Kinderlehrer2000VariationalInequalities}. 
\end{proof}

The existence of CBCP equilibria likewise follows from the variational inequalities in Definition \ref{Def: CBCP Equilibria}.

\begin{proposition} \label{Prop: Lane Flows Existence, CBCP}
For each $(\boldsymbol{\tau}, B)$-CBCP policy, where $\boldsymbol{\tau}$ is component-wise non-negative and $B \geq 0$, there exists a $(\boldsymbol    {\tau}, B)$-CBCP equilibrium.
\end{proposition}


\subsection{Convex Program Characterization}
\label{subsec: Convex Program Characterization}

Although the variational inequalities used to define the DBCP and CBCP equilibria (Definitions \ref{Def: DBCP Equilibria} and \ref{Def: CBCP Equilibria}) naturally imply their existence, they do not provide a computationally tractable mechanism for computing these equilibria. In this section, we address this issue by characterizing any $(\boldsymbol{\tau}, \boldsymbol{\alpha})$-DBCP equilibria as the minimizer of a convex program, and deriving an analogous characterization for $(\boldsymbol{\tau}, B)$-CBCP equilibria under the assumption that the VoTs of eligible users are time-invariant (Remark \ref{Remark: Time-Invariant Eligble User VoTs}).


\begin{theorem} \label{Thm: Convex Program, DBCP} (\textbf{Convex Program for DBCP Equilibria})
The set of feasible flows $\textbf{y}^\star := (\textbf{y}_t: t \in [T]) \in \Y$ is a DBCP Equilibrium if and only if, for each $t \in [T]$, $\textbf{y}_t^\star$ minimizes the following convex program:
{
\begin{align} \label{Eqn: Convex Program Statement, DBCP}
    \min_{\textbf{y} \in \R^{2|\edges||\G|}} \hspace{5mm} &\sum_{e \in \edges} \Bigg[ \sum_{k=1}^2 \int_0^{x_{e,k,t}} \ell_{e,k}(w) \hspace{0.5mm} dw + \sum_{g \in \G_I} \frac{y_{e,1,t}^g \tau_{e,t}}{v_t^g} + \sum_{g \in \G_E} \frac{y_{e,1,t}^g (1 - \alpha_{e,t}) \tau_{e,t}}{v_t^g} \Bigg] \\ \nonumber
    \emph{s.t.} \hspace{5mm} &y_t^g \in \Y_t^{g,d}, \ \forall \ g \in \G, \\ \nonumber
    &\sum_{g \in \G} y_{e,k,t}^g = x_{e,k,t}, \ \forall \ e \in \edges, k \in [2].
\end{align}
}
\end{theorem}

\begin{proof}(\textbf{Sketch})
Theorem \ref{Thm: Convex Program, DBCP} is proved by establishing that, for each time $t \in [T]$, the first-order optimality conditions for the convex program \eqref{Thm: Convex Program, CBCP} correspond precisely to the definition of the corresponding DBCP equilibrium flows (Definition \ref{Def: DBCP Equilibria}). For details, see Appendix \ref{subsec: App A1, Proof of Thm Convex Program, DBCP}.
\end{proof}

The convex program \eqref{Eqn: Convex Program Statement, CBCP} is similar to convex programs that describe equilibrium flows under heterogeneous tolls \cite{Fleischer2004TollsForHeterogeneousSelfishUsers}, though \eqref{Eqn: Convex Program Statement, CBCP} also includes a budget constraint for eligible users. Unlike \eqref{Eqn: Convex Program Statement, DBCP}, the convex program \eqref{Eqn: Convex Program Statement, CBCP} does not decouple across times, since the budget constraint couples eligible users’ travel decisions across the entire time horizon.

\begin{theorem} \label{Thm: Convex Program, CBCP}(\textbf{Convex Program for CBCP Equilibria})
Suppose the value-of-time (VoT) of each eligible group is time-invariant, i.e., for each eligible group $g \in \G_E$, there exists some $v_g > 0$ such that $v_t^g = v_g$ for each $t \in [T]$. Then, the set of feasible flows $\textbf{y}^\star \in \Y^b$ is a CBCP Equilibrium if and only if it minimizes the following convex program:
{
\begin{align} \label{Eqn: Convex Program Statement, CBCP}
    \min_{\textbf{y} \in \R^{3
    |\edges|T|\G|}} \hspace{2mm} &\sum_{t \in [T]} \sum_{e \in \edges} \left[ \sum_{k=1}^2 \int_0^{x_{e,k,t}} \ell_{e,k}(w) \hspace{0.5mm} dw  + \sum_{g \in \G} \frac{\hat y_{e,1,t}^g \tau_{e,t}}{v_t^g} \right] \\ \nonumber
    \text{s.t.} \hspace{5mm} &\textbf{y} \in \Y^b, \\ \nonumber
    &\sum_{g \in \G} y_{e,k,t}^g = x_{e,k,t}, \ \forall \ e \in \edges, k \in [2], t \in [T].
\end{align}
}
\end{theorem}

\begin{proof}(\textbf{Sketch})
Similar to Theorem \ref{Thm: Convex Program, DBCP}, 
Theorem \ref{Thm: Convex Program, CBCP} follows by showing that the first-order optimality conditions to the convex program \eqref{Eqn: Convex Program Statement, CBCP} are equivalent to the definition of the corresponding CBCP equilibrium. 
For details, see Appendix \ref{subsec: App A1, Proof of Thm Convex Program, CBCP}.
\end{proof}

As a result of Theorems \ref{Thm: Convex Program, DBCP} and \ref{Thm: Convex Program, CBCP}, the equilibrium edge flows $x_{e,k,t}^\star$ are unique given any $(\boldsymbol{\tau}, \boldsymbol{\alpha})$-DBCP policy, or any $(\boldsymbol{\tau}, B)$-CBCP policy if all eligible users' VoTs are time-invariant. Unlike \eqref{Eqn: Convex Program Statement, DBCP}, the convex program \eqref{Eqn: Convex Program Statement, CBCP} does not decouple across times, since the budget constraint couples eligible users’ travel decisions across the entire time horizon.

\begin{remark}
The convex program \eqref{Eqn: Convex Program Statement, CBCP} is similar to convex programs that describe equilibrium flows under heterogeneous tolls \cite{Fleischer2004TollsForHeterogeneousSelfishUsers}, though \eqref{Eqn: Convex Program Statement, CBCP} also includes a budget constraint for eligible users.
\end{remark}

\begin{remark} \label{Remark: Time-Invariant Eligble User VoTs}
Whereas the convex program \eqref{Eqn: Convex Program Statement, DBCP} in Theorem \ref{Thm: Convex Program, DBCP} can be used to compute any $(\boldsymbol{\tau}, \boldsymbol{\alpha})$-DBCP equilibria, the convex program \eqref{Eqn: Convex Program Statement, CBCP} in Theorem \ref{Thm: Convex Program, CBCP} can be used to compute $(\boldsymbol{\tau}, B)$-CBCP equilibria only under the setting where each eligible user group has time-invariant VoTs. On a technical level, this assumption allows the first-order optimality conditions of the convex program \eqref{Eqn: Convex Program Statement, CBCP} to match the definition of the corresponding CBCP equilibrium. On a practical level, as noted in Jalota et al. \cite{Jalota2023CreditBasedCongestionPricing}, users' VoTs for their commute during work days are unlikely to change significantly over time. More details on the real-world implications of this assumption are provided in Jalota et al. \cite{Jalota2023CreditBasedCongestionPricing}.
\end{remark}

\subsection{Uniqueness of Edge Flows}
\label{subsec: Uniqueness of Edge Flows}


The convex programs \eqref{Thm: Convex Program, DBCP} and \eqref{Thm: Convex Program, CBCP} in Section \ref{subsec: Convex Program Characterization} used to characterize DBCP and CBCP equilibria, respectively, imply the uniqueness of the aggregate lane flows at equilibria.

\begin{proposition}[\textbf{Uniqueness of Lane Flows at Equilibrium, DBCP}] \label{Prop: Lane Flows Uniqueness, DBCP}
For any DBCP scheme $(\boldsymbol{\tau}, \boldsymbol{\alpha})$, the aggregate edge flow $x \in \R^{2|\edges|T}$ corresponding to any CBCP $(\boldsymbol{\tau}, \boldsymbol{\alpha})$-equilibrium is unique.
\end{proposition}

\begin{proof}
See Appendix \ref{subsec: App A1, Proof of Prop Equilibrium Lane Flows Uniqueness, DBCP}.
\end{proof}

\begin{proposition}[\textbf{Uniqueness of Edge Flows at Equilibrium, CBCP}] \label{Prop: Lane Flows Uniqueness, CBCP}
Suppose the value-of-time (VoT) of each eligible group is time-invariant, i.e., for each eligible group $g \in \G_E$, there exists some $v_g > 0$ such that $v_t^g = v_g$ for each $t \in [T]$. Then for any CBCP scheme $(\boldsymbol{\tau}, B)$, the aggregate edge flow $x \in \R^{3|\edges|T}$ corresponding to any CBCP $(\boldsymbol{\tau}, B)$-equilibrium is unique.
\end{proposition}

\begin{proof}
See Appendix \ref{subsec: App A1, Proof of Prop Equilibrium Lane Flows Uniqueness, CBCP}.
\end{proof}


\section{Budget vs Discount}
\label{sec: Budget vs Discount}


In this section, we compare the eligible users' equilibrium flow patterns under DBCP and CBCP policies that offer the same travel credit, either as a lump sum or a toll discount. 
\revision{To address the issue that conventional congestion pricing policies often deploy tolls that price low-income users out of accessing express lanes, we compare eligible users' express lane use under CBCP and DBCP policies to study the relative efficacy of CBCP and DBCP policies in promoting equitable express lane usage.}
Studying eligible users' occupancy of the express lane enables us to compute users' equilibrium travel costs, and thus compare the benefits of deploying CBCP versus DBCP policies given a societal welfare objective. 
In Sec. \ref{subsec: Setting}, we introduce new assumptions that allow us to derive key insights regarding the performance of DBCP and CBCP policies, while capturing salient aspects of current, real-world DBCP and CBCP policies. In Sec. \ref{subsec: Case 1: 1 el}, we compare the eligible users' express lane flows at equilibrium under various CBCP and DBCP policies. In Sec. \ref{subsec: Case 2: 1 el, 1 in}, we repeat this analysis for the case where the user population consists of one eligible group and one ineligible group.
\revision{Both Sec. \ref{subsec: Case 1: 1 el} and \ref{subsec: Case 2: 1 el, 1 in} specify regimes of DBCP and CBCP equilibria, parameterized by the toll subsidy provided to eligible users, in which DBCP policies outperform CBCP policies in facilitating eligible users' access to tolled express lanes, or vice versa. We also pinpoint user-dependent and network-specific parameters, such as users' VoTs and the latency function, that play a crucial role in deciding whether CBCP or DBCP policies are more successful in promoting eligible users' express lane usage. Our analysis demonstrates the importance of acquiring accurate knowledge of user and network attributes when designing congestion pricing policies.
}

\subsection{Setting}
\label{subsec: Setting}

We assume in this section that the traffic network $\network$ consists of a single edge with an express and a general purpose lane, as in Jalota et al. \cite{Jalota2023CreditBasedCongestionPricing}, and that the time horizon $T$ is 1. Thus, the toll and discount are scalars, denoted $\tau$ and $\alpha$ below, respectively. Since current tolled lanes are usually implemented on isolated highway segments, equilibrium analyses of DBCP and CBCP policies over single-edge models describe user flow patterns on tolled express lanes 
well. Also, results derived under the single time step assumption extend naturally to arbitrary time horizons when the tolling policy considered is time-invariant (which is common in practice), in which case eligible users will split their budget use equally across the time horizon.

We also make the following assumption.

\begin{assumption} \label{Assumption: Latency Function} 
The express and general purpose lanes share the same latency function $\ell$, which has strictly positive third derivative in its domain.
\end{assumption}

In practice, express and general purpose lanes often share latency functions, as they describe different lanes on the same road segment. In Sec. \ref{subsec: Setup}, we already assumed that $\ell$ is strictly positive, strictly increasing, and strictly convex. Here, for technical reasons, we additionally assume that $\ell$ has strictly positive third derivatives, which holds for most real-world latency functions
\cite{BPR1964TrafficAssignmentManual}. Sec. \ref{sec: Experimental Results} includes empirical results that compare DBCP and CBCP equilibria
in settings in which the above assumptions are relaxed.

\begin{remark} \label{Remark: CBCP and DBCP, Fair Comparison}
We aim to contrast the steady-state effects of providing the same travel subsidy as a lump sum versus as a toll discount. Under $(\tau, B)$-CBCP policies, each eligible group is allotted at most $B$ in travel credit over the time horizon $T$,  whereas under $(\tau, \alpha)$-DBCP policies, each eligible group is allotted at most $\alpha \tau$ in travel assistance. Thus, to compare budget-based and discount-based travel assistance policies, below we contrast the outcomes of deploying the $(\tau, \alpha)$-DBCP policy versus the $(\tau, \alpha \tau)$-CBCP policy. 
\end{remark}


In Sec. \ref{subsec: Case 1: 1 el} and \ref{subsec: Case 2: 1 el, 1 in}, let $y^C(\alpha)$ and $y^D(\alpha)$ denote the eligible user's express lane flow at the $(\tau, \alpha \tau)$-CBCP and $(\tau, \alpha)$-DBCP equilibria, respectively, for any $\alpha \in (0, 1)$ at which these equilibrium flows are well-defined. We compare $y^C(\alpha)$ and $y^D(\alpha)$ as $\alpha$ is varied from 0 to 1, to decide, at each value of $\alpha$, whether CBCP or DBCP policies are more effective in improving eligible users' express lane access. 

\subsection{Case 1: Single Eligible Group}
\label{subsec: Case 1: 1 el}

Here, we study the setting where the entire user population is eligible for travel subsidies, and show that DBCP policies outperform CBCP policies in promoting eligible users' express lane access when the discount level $\alpha$ is sufficiently small. We begin by making the following assumption.

\begin{assumption} \label{Assumption: For 1 eligible group, 0 ineligible group}
Suppose that:
\begin{enumerate}
    \item The user population consists of one eligible group with time-invariant VoT $v^E$.

    \item $\tau < v^E (\ell(1) - \ell(0))$.
\end{enumerate}
\end{assumption}

As discussed in Sec. \ref{sec: Equilibria Characterization}, user groups' VoTs are well-modeled as time-invariant.
Moreover, we study the setting when $\tau < v^E (\ell(1) - \ell(0))$, since otherwise, the toll would be so high that under CBCP policies, an eligible user would never pay out of pocket to access the express lane.



\begin{figure}
    \centering
    \includegraphics[scale=0.15]{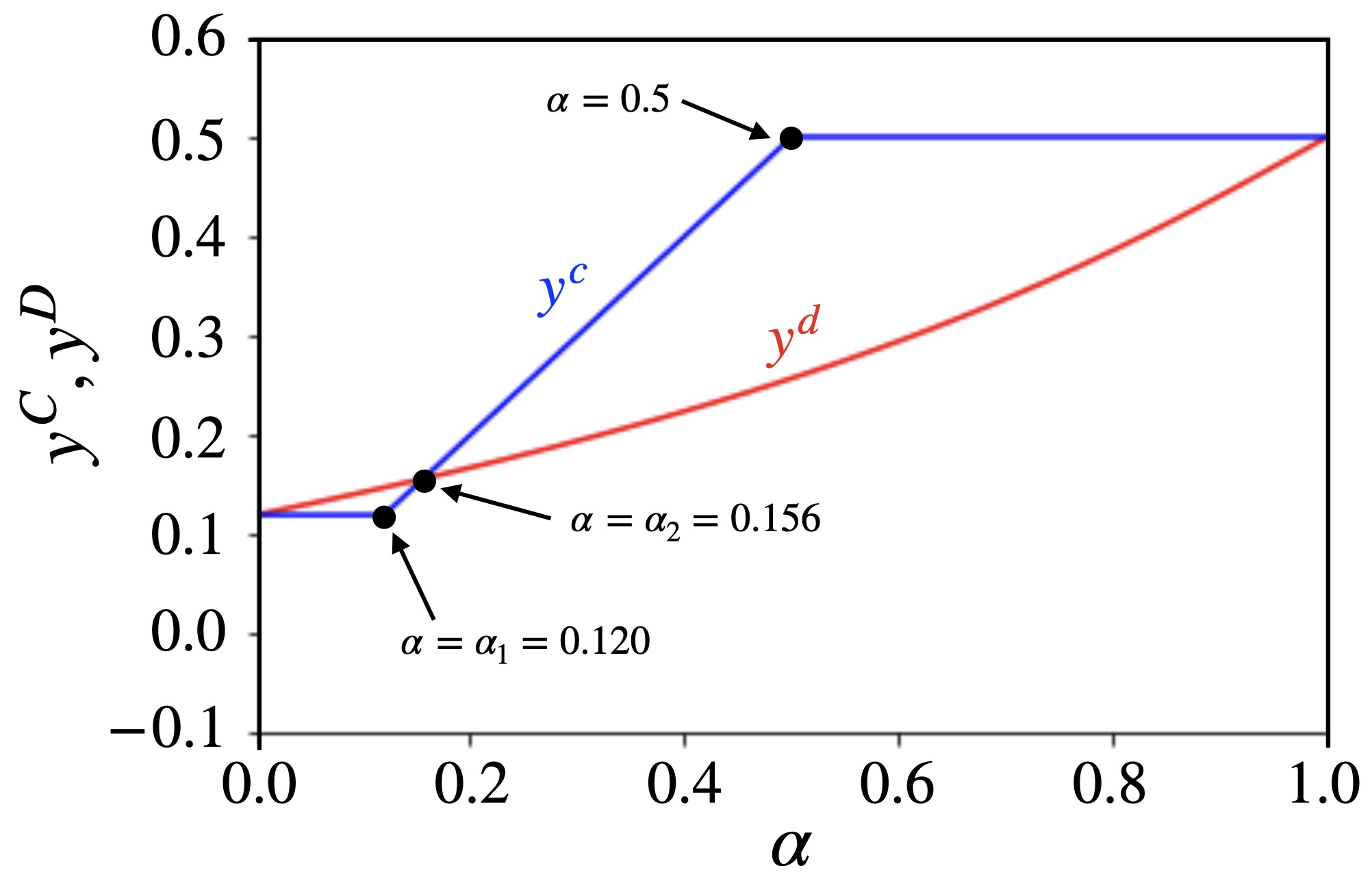}
    \caption{$y^C$ vs. $\alpha$ and $y^D$ vs. $\alpha$, under Assumptions \ref{Assumption: Latency Function} and \ref{Assumption: For 1 eligible group, 0 ineligible group}, for the setting where $\ell(x) = x^4/16$, $\tau = 0.6$, and $v^E = 1$, in which case $\alpha_1 = 0.120$, $\alpha_2 = 0.156$.}
    \label{fig:y_vs_alpha_10}
\end{figure}

\begin{figure}
    \centering
    \includegraphics[scale=0.17]{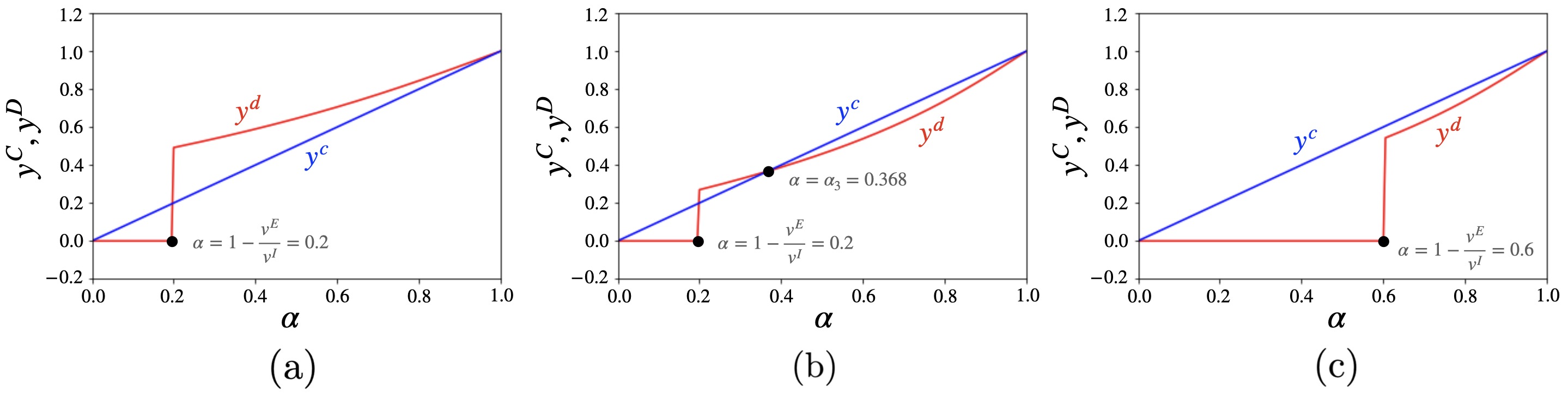}
    \caption{$y^C(\alpha)$ and $y^D(\alpha)$, under Assumptions \ref{Assumption: Latency Function} and \ref{Assumption: For 1 eligible group, 1 ineligible group}, for the settings where $\ell(x) = x^4/16$ and (Left) $\tau = 0.4$, $v^E = 1$, $v^I = 1.25$, in which case $\tau < 2v^E \ell'(1)$, (Middle) $\tau = 0.7$, $v^E = 1$, $v^I = 1.25$, in which case $\tau > 2v^E \ell'(1)$, $1 - v^E/v^I = 0.2 < \alpha_3 = 0.368$, (Right) $\tau = 0.7$, $v^E = 1$, $v^I = 2.5$, in which case $\tau > 2v^E \ell'(1)$, $1 - v^E/v^I = 0.6 > \alpha_3 = 0.368$.}
    \label{fig:y_vs_alpha_11}
\end{figure}

We begin by computing $y^C(\alpha)$ and $y^D(\alpha)$ under Assumptions \ref{Assumption: Latency Function} and \ref{Assumption: For 1 eligible group, 0 ineligible group} (Theorem \ref{Thm: yc, yd, 10}). Figure \ref{fig:y_vs_alpha_10} presents a plot of $y^C(\alpha)$ and $y^D(\alpha)$ for a particular set of latency function, toll, eligible users' VoT, and discount values.

\begin{theorem} \label{Thm: yc, yd, 10}
Under Assumptions \ref{Assumption: Latency Function} and \ref{Assumption: For 1 eligible group, 0 ineligible group}, there exists a unique $\alpha_1 \in (0, 1/2)$ such that $v^E \ell(\alpha_1) + \tau = v^E \ell(1 - \alpha_1)$. Then, for each $\alpha \in (0, 1)$, the $(\tau, \alpha \tau)$-CBCP equilibrium express lane flow $y^C(\alpha)$ is given by:
\begin{align} \label{Eqn: yc, 10}
    y^C(\alpha) &:= \begin{cases}
        \alpha_1, \hspace{5mm} &\alpha \in (0, \alpha_1), \\
        \alpha, \hspace{5mm} &\alpha \in (0, 1/2), \\
        1/2, \hspace{5mm} &\alpha \in (1/2, 0).
    \end{cases}
\end{align}
while the $(\tau, \alpha)$-DBCP equilibrium express lane flow $y^D(\alpha)$ is given by the unique solution to the fixed-point equation $v^E \ell(y^D(\alpha)) + (1-\alpha) \tau = v^E \ell(y^D(1 - \alpha))$.
\end{theorem}

\begin{proof}
See Appendix \ref{subsec: App A2, Proof of Thm yc, yd, 10}.
\end{proof}

In words, as $\alpha$ increases from $0$ to $1$, the express lane flow at the $(\tau, \alpha)$-DBCP equilibrium, i.e., $y^D(\alpha)$, gradually (but strictly) increases. Meanwhile, the express lane flow at the $(\tau, \alpha \tau)$-CBCP equilibrium, i.e., $y^C(\alpha)$, stays constant until $\alpha$ exceeds $\alpha_1$, at which point $y^C(\alpha)$ begins rising more rapidly. Finally, after $\alpha$ reaches $1/2$, both lanes become equally crowded, and $y^C(\alpha)$ from then on is fixed at $1/2$.

\revision{Given a toll level $\tau$, the following theorem establishes that when all users are eligible, there exists a critical discount level $\alpha_2$ such that the $(\tau, \alpha)$-DBCP policy promotes eligible users' express lane access to a greater extent compared to the $(\tau, B)$-CBCP policy if and only if $\alpha > \alpha_2$. From a policy perspective, our theoretical analysis underscores the importance of accurately estimating users' VoTs and the latency function $\ell$, both of which affect the value of $\alpha_2$, before designing traffic regulation mechanisms.
}

\begin{theorem} \label{Thm: yc vs. yd, 10}
Under Assumptions \ref{Assumption: Latency Function} and \ref{Assumption: For 1 eligible group, 0 ineligible group}, $y^D(\alpha)$ is strictly increasing and strictly convex. Thus, there is a unique $\alpha_2 \in (\alpha_1, 1/2)$ such that $v^E \ell(\alpha_2) + (1 - \alpha_2) \tau = v^E \ell(1 - \alpha_2)$. Moreover, $y^D(\alpha) > y^C(\alpha)$ for each $\alpha \in (0, \alpha_2)$, and $y^D(\alpha) < y^C(\alpha)$ for each $\alpha \in (\alpha_2, 1)$.
\end{theorem}

\begin{proof}
See Appendix \ref{subsec: App A2, Proof of Thm yc vs. yd, 10}.
\end{proof}

As Figure \ref{fig:y_vs_alpha_10} shows, $y^C(\alpha)$ is a (weakly) increasing function with two constant segments: $y^C(\alpha) = \alpha_1, \forall \alpha \in (0, \alpha_1)$ and $y^C(\alpha) = 1/2, \forall \alpha \in (1/2, 1)$. Meanwhile, $y^D(\alpha)$ strictly increases from $\alpha_1$ to $1/2$ when $\alpha$ increases from 0 to 1. Thus, 
there exists some $\alpha_2 \in (0, 1)$ at which $y^C(\alpha_2) = y^D(\alpha_2)$. Moreover, as $y^D(\alpha)$ is strictly convex, $\alpha_2$ is unique.


\subsection{Case 2: Single Eligible and Single Ineligible Group}
\label{subsec: Case 2: 1 el, 1 in}

Here, we study the setting where the user population consists of one eligible and one ineligible user group, each of which has demand equal to one. We show that in this setting, the relative performance of DBCP and CBCP policies, in terms of promoting eligible users' express lane access, depends not only on the discount level $\alpha$, but also on the toll $\tau$, ineligible users' VoT $v^I$, and latency function $\ell$. \revision{Thus, our analysis emphasizes the need for accurate estimates of users' VoTs and the latency function $\ell$ when designing congestion pricing mechanisms.}

First, we give the analog of Assumption \ref{Assumption: For 1 eligible group, 0 ineligible group} in this setting.

\begin{assumption} \label{Assumption: For 1 eligible group, 1 ineligible group}
Suppose that:
\begin{enumerate}
    \item The user population consists of one eligible and one ineligible group, with respective time-invariant VoTs $v^E$ and $v^I$. Moreover, $v^E < v^I$.

    \item $\tau < v^E (\ell(2) - \ell(0))$.
\end{enumerate}
\end{assumption}

The assumption $v^E < v^I$ reflects that eligible users often have less income than ineligible users, and are thus associated with lower VoTs, as evinced by real-world census data \cite{BPR1964TrafficAssignmentManual}. Our analysis generalizes to the $v^E \geq v^I$ setting. 


We compute $y^C(\alpha)$ and $y^D(\alpha)$ under Assumptions \ref{Assumption: Latency Function} and \ref{Assumption: For 1 eligible group, 1 ineligible group} (Theorem \ref{Thm: yc, yd, 11}), and explain our findings. Figure \ref{fig:y_vs_alpha_11} plots $y^C(\alpha)$ and $y^D(\alpha)$ for particular choices of latency functions, tolls, eligible users' VoTs, and discount values.

\begin{theorem} \label{Thm: yc, yd, 11}
Under Assumptions \ref{Assumption: Latency Function} and \ref{Assumption: For 1 eligible group, 1 ineligible group}:
\begin{enumerate}
    \item $\forall \alpha \in [0, 1],$ we have $y^C(\alpha) = \alpha.$
    
    \item $\forall \alpha \in (0, 1 - v^E/v^I)$, we have $y^D(\alpha) = 0$, and $\forall \alpha \in (1 - v^E/v^I, 1]$, $y^D(\alpha)$ is the unique solution to the fixed-point equation $v^E \ell(y^D(\alpha)) + (1-\alpha) \tau = v^E \ell(y^D(2 - \alpha))$, and is strictly increasing and strictly convex.
\end{enumerate}
\end{theorem}

\begin{proof}
See Appendix \ref{subsec: App A2, Proof of Thm yc, yd, 11}.
\end{proof}

When a $(\tau, \alpha \tau)$-CBCP policy is deployed, the eligible users’ equilibrium express lane flow associated with credit-based toll payments is $\alpha$, i.e., $\tilde y_1 = \alpha$.
No eligible user pays the toll out of pocket, i.e., $\hat y_1 = 0$, as they are priced out by ineligible users who have higher VoTs.
Thus, $y^C(\alpha) = \alpha, \forall \alpha = (0, 1)$. Similarly, when a $(\tau, \alpha)$-DBCP policy is deployed, if $\alpha < 1 - v^E / v^I$, the toll discount provided is insufficient to prevent eligible users from being priced out of express lane access, so $y^D(\alpha) = 0$. However, after $\alpha$ exceeds $1 - v^E/v^I$, the eligible and ineligible users' roles reverse, and $y^D(\alpha)$ jumps upward to the solution of the fixed-point equation $v^E \ell(y^D(\alpha)) + (1 - \alpha) \tau = v^E \ell(y^D(2 - \alpha))$. 

\begin{theorem} \label{Thm: yc vs. yd, 11}
There exists a unique $\alpha_3 \in (0, 1)$ such that $v^E \ell(\alpha_3) + (1 - \alpha_3) \tau = v^E \ell(1 - \alpha_3)$. Moreover, let $\ell'$ denote the derivative of the latency function $\ell$. Then:
\begin{enumerate}
    \item If $\tau < 2v^E \ell'(1)$, then $y^D(\alpha) < y^C(\alpha) \ \forall \alpha \in \big(0, 1 - \frac{v^E}{v^I} \big)$, and $y^D(\alpha) > y^C(\alpha) \ \forall \alpha \in \big(1 - \frac{v^E}{v^I}, 1 \big)$.

    \item If $\tau > 2v^E \ell'(1)$ and $\frac{v^E}{v^I} > 1 - \alpha_3$, then $y^D(\alpha) < y^C(\alpha) \ \forall \alpha \in \big(0, 1 - \frac{v^E}{v^I} \big) \cup (\alpha_3, 1)$, and $y^D(\alpha) > y^C(\alpha) \ \forall \alpha \in \big(1 - \frac{v^E}{v^I}, \alpha_3 \big)$.
    
    \item If $\tau > 2v^E \ell'(1)$ and $\frac{v^E}{v^I} > 1 - \alpha_3$, then $y^D(\alpha) < y^C(\alpha) \ \forall \alpha \in (0, 1)$.
\end{enumerate}
\end{theorem}

\begin{proof}
See Appendix \ref{subsec: App A2, Proof of Thm yc vs. yd, 11}.
\end{proof}

In words, low tolls and high discounts can prevent eligible users from being priced out of express lane access, causing the eligible users' express lane flow to be higher under the corresponding DBCP than CBCP equilibrium.

\section{Experimental Results}
\label{sec: Experimental Results}

Below, we present simulation results that validate the theoretical contributions in Sec. \ref{sec: Budget vs Discount} and explore their sensitivity to ineligible users' demand and VoTs. Sec. \ref{subsec: Sensitivity Analysis} presents sensitivity analysis on properties of DBCP and CBCP equilibria given in Sec. \ref{sec: Budget vs Discount}. \revision
{Specifically, Theorems \ref{Thm: yc vs. yd, 10} and \ref{Thm: yc vs. yd, 11} establish distinct regimes of DBCP and CBCP equilibria in which either DBCP or CBCP policies outperform the other in terms of improving eligible users' express lane access. Our simulation study in \ref{subsec: Sensitivity Analysis} shows that the findings in Theorems \ref{Thm: yc vs. yd, 10} and \ref{Thm: yc vs. yd, 11} are robust with respect to the users' VoTs and the total traffic demand.
} Sec. \ref{subsec: Case Study: Optimal DBCP and CBCP Policies} presents a real-world study of the effects of deploying DBCP and CBCP policies on a highway segment in San Mateo County, California.

\subsection{Sensitivity Analysis}
\label{subsec: Sensitivity Analysis}

We analyze the sensitivity of 
Theorem \ref{Thm: yc vs. yd, 11} by studying variations in eligible users' equilibrium express lane usage under relaxations of Assumption \ref{Assumption: For 1 eligible group, 1 ineligible group}, such as changes in the ineligible users' demand and VoTs. As in Sec. \ref{subsec: Case 2: 1 el, 1 in}, we focus on the setting with a single-edge traffic network shared by an eligible and an ineligible user group. In our experiments, we set the latency function to be $\ell(x) = x^2/4$, a fixed toll of $\tau = 0.4$, and eligible users' VoT at $v^E = 1.0$.

\begin{figure}
    \centering
    \includegraphics[scale=0.15]{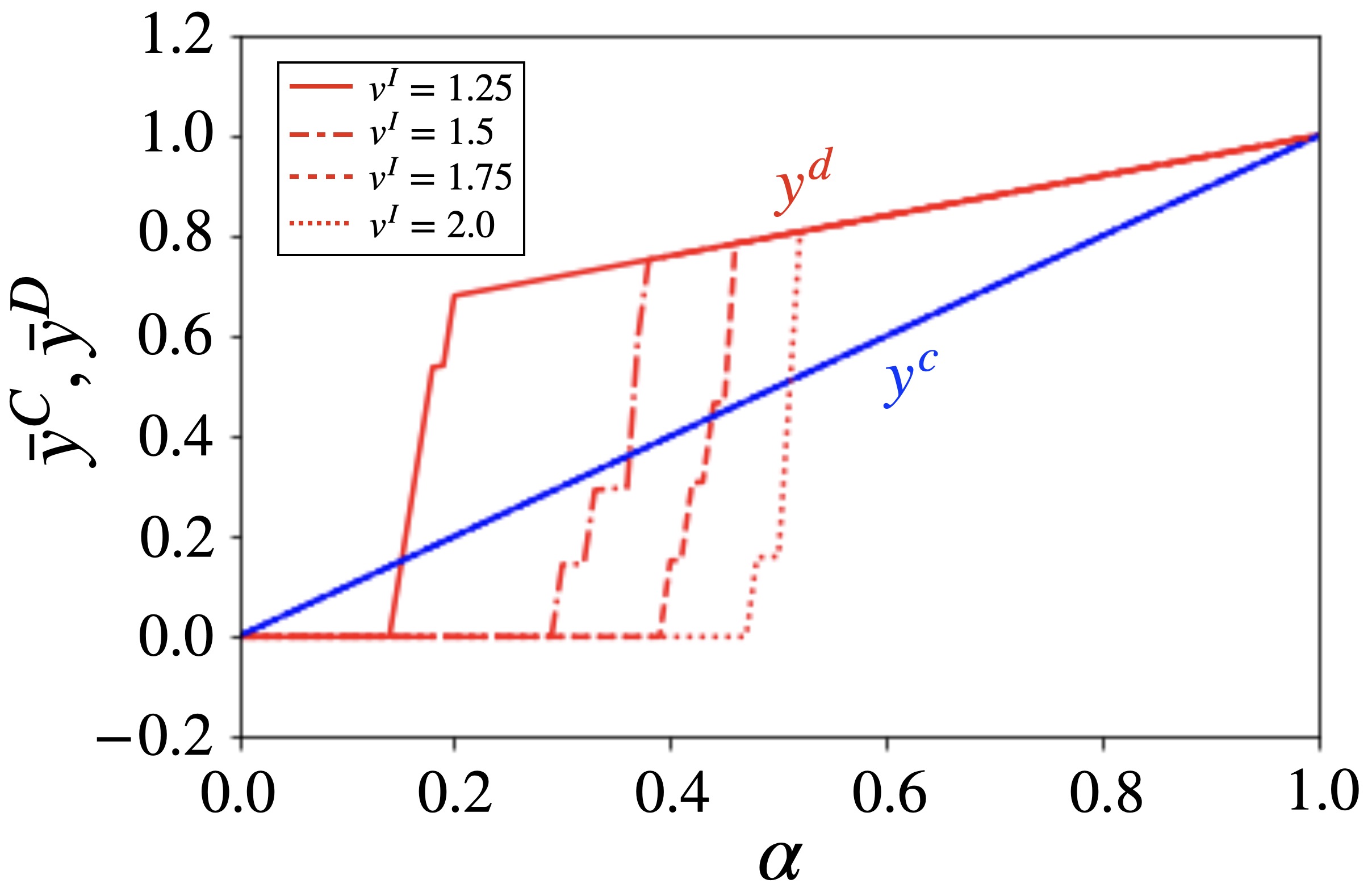}
    \caption{$\bar y^C$ (red) and $\bar y^D$ (blue) vs. $\alpha$ at $v^I = 1.25, 1.5, 1.75, 2.0$.}
    \label{fig:sensitivity_VoT}
\end{figure}


First, we study variations in eligible users' express lane usage when the ineligible users' VoT $v^I$ changes over time. We extend the time horizon to $T = 5$ and set the ineligible users' VoT at each time to be $v^I = \bar v^I + u \Delta v^I$, where $\bar v^I$ is a fixed baseline value for the ineligible users' VoT, $u$ is drawn uniformly from $[-1, 1]$, and $\Delta v^I$ captures the perturbation of $v^I$ at each time. We compute, for each $\alpha \in [0, 1]$, the eligible users' average express lane flow over the time horizon $T$ at $(\tau, \alpha \tau)$-CBCP and $(\tau, \alpha)$-DBCP equilibria, denoted $\bar y^C(\alpha)$ and $\bar y^D(\alpha)$, respectively. Figure \ref{fig:sensitivity_VoT} plots $\bar y^C(\alpha)$ and $\bar y^D(\alpha)$ at $v^I = 1.25, 1.5, 1.75, 2.0$ and $\Delta v^I = 0.1$. At each level of $\bar v^I$, $y^D(\alpha)$ slightly fluctuates under DBCP policies when $v^I$ changes over time. However, $y^C(\alpha)$ is unaffected, as CBCP policies guarantee eligible users a degree of access to express lanes proportional to the budget allotted, irrespective of $v^I$.

\begin{figure}
    \centering
    \includegraphics[scale=0.15]{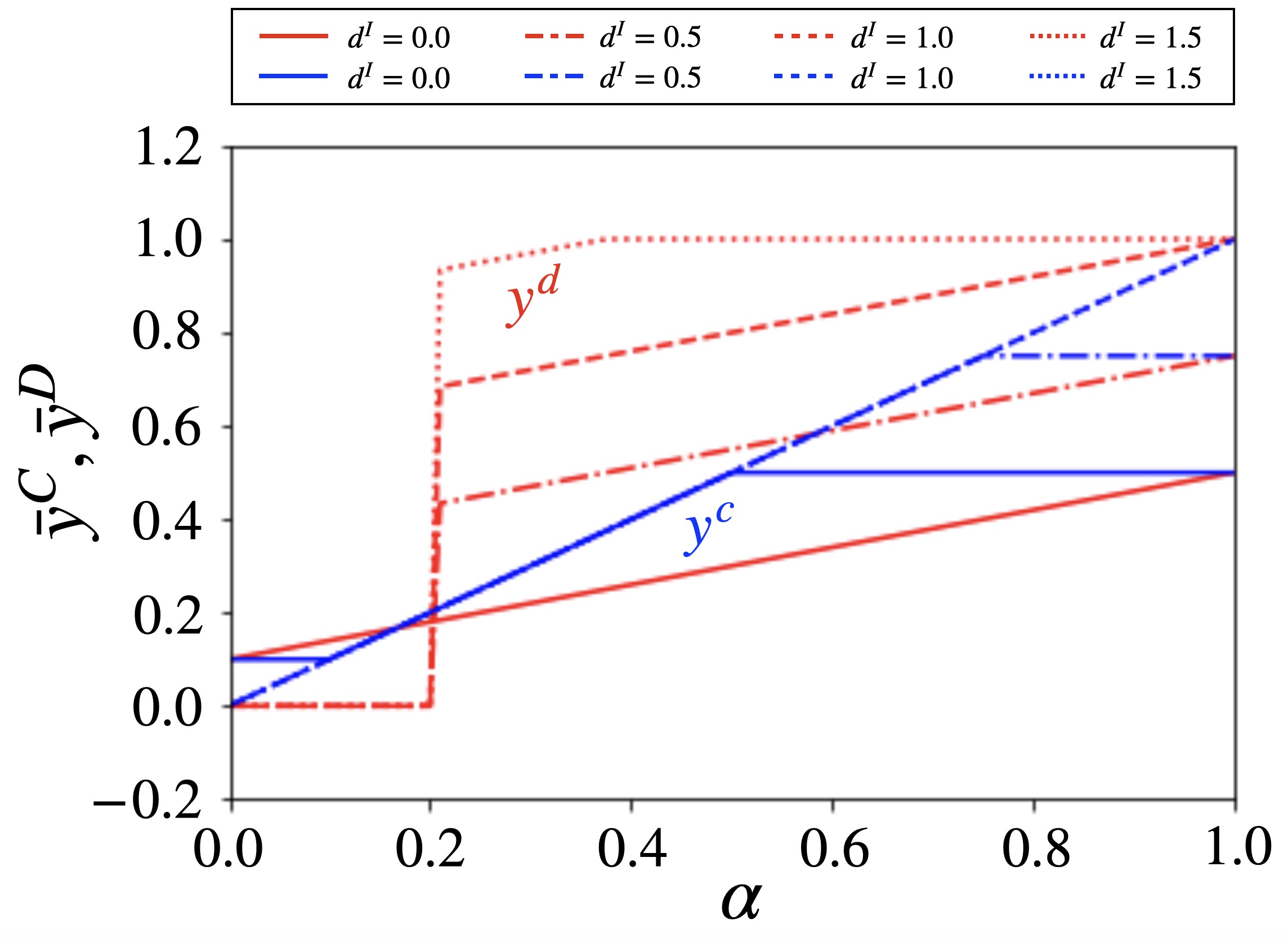}
    \caption{$y^C$ (red) and $y^D$ (blue) vs. $\alpha$, at $d^I = 0, 0.5, 1, 1.5$.}
    \label{fig:sensitivity_demand}
\end{figure}

Next, we describe changes in the eligible users' express lane use when ineligible users' demand, denoted $d^I$ below, varies. We again set $T = 1$ and compute, for each $\alpha \in [0, 1]$, the eligible users' express lane flow at $(\tau, \alpha\tau)$-CBCP and $(\tau, \alpha)$-DBCP equilibria.
Figure \ref{fig:sensitivity_demand} plots $y^C(\alpha)$ and $y^D(\alpha)$ at $d^I = 0.0, 0.5, 1.0, 1.5$. As $d^I$ varies from $0$ to $1$, $y^C(\alpha)$ and $y^D(\alpha)$ interpolate the equilibrium flows for the setting with a single eligible user group (Sec. \ref{subsec: Case 1: 1 el}), which corresponds to the $d^I = 0$ setting, and with one eligible and one ineligible group (Sec. \ref{subsec: Case 2: 1 el, 1 in}), which corresponds to the $d^I = 1$ setting.

\subsection{Case Study: Optimal DBCP and CBCP Policies}
\label{subsec: Case Study: Optimal DBCP and CBCP Policies}

Here, we explore the impact of deploying CBCP and DBCP policies on a four-lane highway (one express lane, three general purpose lanes), within the framework of the San Mateo 101 Express Lanes Project. As in Jalota et al. \cite{Jalota2023CreditBasedCongestionPricing}, we use data from the Caltrans Performance Measurement System (PeMS) database \cite{pems-database} to estimate the BPR latency function \cite{BPR1964TrafficAssignmentManual}, and income data from the 2020 US Census American Community Survey (ACS) to estimate all users' VoTs. A detailed description of 
these model parameters is given in Jalota et al. \cite{Jalota2023CreditBasedCongestionPricing}. We sample tolls from \$0 to \$20, at \$1 increments, and budgets from \$0 to \$90, at \$5 increments. We then solve the convex programs \eqref{Eqn: Convex Program Statement, DBCP} and \eqref{Eqn: Convex Program Statement, CBCP} to compute the $(\boldsymbol{\tau}, B)$-CBCP and $(\boldsymbol{\tau}, \alpha)$-DBCP equilibrium flows at each toll and budget level, with $\alpha = B/(\tau T)$ to ensure a fair comparison between the two types of policies (Remark \ref{Remark: CBCP and DBCP, Fair Comparison}). We describe variations in equilibrium express lane flow levels and average travel times across different CBCP and DBCP policies (Sec. \ref{subsubsec: Express Lane Flows and Average Travel Times}), and sample $\tau$ and $B$ to optimize societal welfare measures (Sec. \ref{subsubsec: Optimizing Societal Cost}).

\subsubsection{Express Lane Flows and Average Travel Times}
\label{subsubsec: Express Lane Flows and Average Travel Times}

\begin{figure}
    \centering
    \includegraphics[scale=0.25]{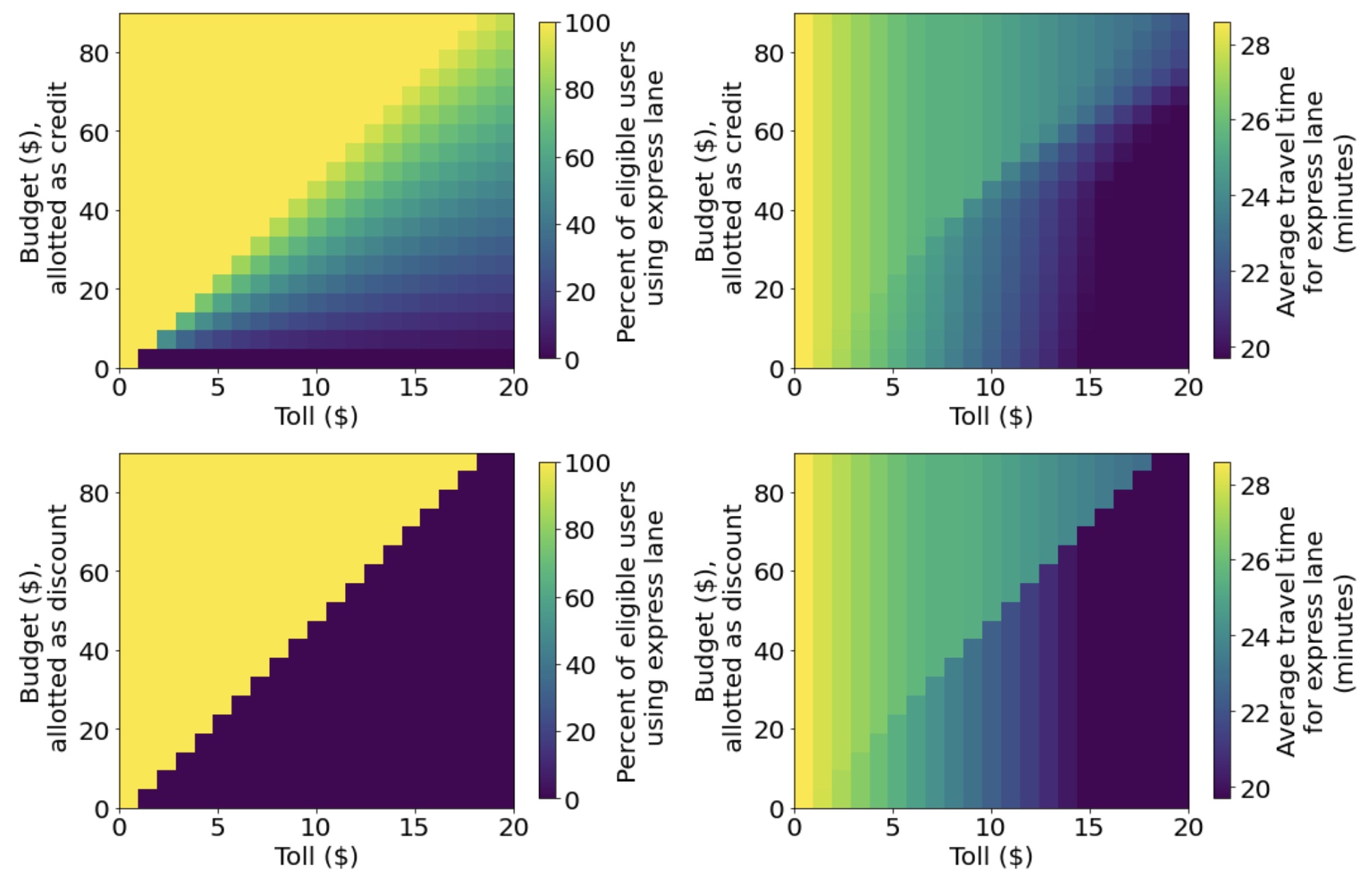}
    \caption{(Left) Percent of eligible users and (Right) Average eligible users' travel time at (Top) $(\tau, B)$-CBCP and (Bottom) $(\tau, \alpha = B/(\tau T))$-DBCP equilibria with toll $\tau$ and allotted credit $B$.}
    \label{fig:San_Mateo_color_plots}
\end{figure}

Figure \ref{fig:San_Mateo_color_plots} plots eligible users' express lane usage and average travel times at $(\boldsymbol{\tau}, B)$-CBCP and $(\boldsymbol{\tau}, \alpha)$-DBCP equilibria, with $\alpha = B/(\tau T)$, across the range of tolls and budgets described above. We make the following observations, which corroborate the theory in Sec. \ref{subsec: Case 2: 1 el, 1 in}. Under both CBCP and DBCP policies, the fraction of eligible users on the express lane at equilibrium rises with the amount of travel subsidy provided, and drops when the toll rises. The average express lane travel time shifts in the opposite direction, since higher traffic flows induce higher travel times on the express lane. In accordance with Theorem \ref{Thm: yc vs. yd, 11}, variations in the toll or budget cause more sudden shifts in eligible users' express lane usage under equilibrium flows for discount-based policies compared to budget-based policies.

\subsubsection{Optimizing Societal Cost}
\label{subsubsec: Optimizing Societal Cost}

Here, our objective is to minimize the following measure of societal cost:
{ 
\begin{align*}
    f_{\boldsymbol{\lambda}} &:= \lambda_E \cdot \sum_{g \in \G_E} \sum_{t \in [T]} \Big( \tau_t \hat y_{1,t}^g + \sum_{k \in [2]} v_g \ell_k(x_{k,t}) y_{k,t}^g \Big) + \lambda_I \cdot \sum_{g \in \G_I} \sum_{t \in [T]} \Big( \tau_t \hat y_{1,t}^g + \sum_{k \in [2]} v_g \ell_k(x_{k,t}) y_{k,t}^g \Big) 
    \\
    &\hspace{5mm} 
    - \lambda_R \cdot \sum_{g \in \G} \sum_{t \in [T]} \tau_t \hat y_{1,t}^g.
\end{align*}
}
Here, the Pareto weights $\boldsymbol{\lambda} := (\lambda_E, \lambda_I, \lambda_R)$ capture the relevant importance of the eligible user travel and toll costs, ineligible users' travel and toll costs, and (negative) toll revenue, respectively. Given any $\boldsymbol{\lambda}$, we aim to find the optimal $(\boldsymbol{\tau}, B)$-CBCP and equivalent $(\boldsymbol{\tau}, \alpha)$-DBCP policies (i.e., $\alpha = B/(\tau T)$) whose equilibrium flows minimize $f_{\boldsymbol{\lambda}}$. 
Table \ref{table:ParetoResults} gives toll and budget values whose 
corresponding DBCP and CBCP
equilibrium flows minimize $f_{\boldsymbol{\lambda}}$, for a range of weights $\boldsymbol{\lambda}$. As noted in Jalota et al. \cite{Jalota2023CreditBasedCongestionPricing}, the optimal toll and budget depend on $\boldsymbol{\lambda}$. Whether the optimal CBCP or DBCP policy achieves a lower minimum cost $f_\lambda$ also depends on  $\boldsymbol{\lambda}$. For example, the optimal CBCP policy achieves a lower value of $f_\lambda$ when only the eligible users' travel and toll costs are considered ($\boldsymbol{\lambda} = (1, 0, 0)$), but a higher value when the toll revenue is also taken into account ($\boldsymbol{\lambda} = (1, 1, 0)$). These variations underscore the importance of systematically analyzing DBCP and CBCP policies at all toll and budget values, since either the optimal DBCP or CBCP policy could outperform the other given a specific societal objective.


\begin{table}[t]
\centering
\begin{tabular}{c|cc|ccc|cc}
\toprule
        \multicolumn{1}{c|}{Weights} & \multicolumn{2}{c|}{Optimal CBCP} & \multicolumn{3}{c|}{\% using express lane} & \multicolumn{2}{c}{Average TT}  \\
        ($\lambda_E$,$\lambda_I$,$\lambda_R$) &$\tau$ & $B$ &Overall  & Eligible  &Ineligible&Express&GPL\\
        \midrule
        (1, 0, 0) & 20 & 90 & 18.5 & 90.0 & 3.8 & 21.4 & 26.1 \\
        (0, 1, 0) & 7 & 35 & 21.7 & 100.0 & 5.7 & 24.7 & 25.6 \\
        (0, 0, 1) & 15 & 0 & 15.7 & 0.0 & 18.9 & 20.2 & 26.6 \\
        (10, 1, 1) & 11 & 0 & 18.3 & 0.0 & 22.0 & 21.3 & 26.2 \\
        (11, 1, 1) & 13 & 45 & 18.1 & 69.2 & 7.6 & 21.1 & 26.2 \\
        (15, 1, 1) & 19 & 75 & 18.1 & 78.9 & 5.7 & 21.2 & 26.2 \\
        \midrule
        \multicolumn{1}{c|}{Weights} & \multicolumn{2}{c|}{Optimal DBCP} & \multicolumn{3}{c|}{\% using express lane} & \multicolumn{2}{c}{Average TT}  \\
        ($\lambda_E$,$\lambda_I$,$\lambda_R$) &$\tau$ & $B (= \alpha T \tau)$ &Overall  & Eligible  &Ineligible&Express&GPL\\
        \midrule
        (1, 0, 0) & 18 & 90 & 19.4 & 100.0 & 2.9 & 22.1 & 26.0 \\
        (0, 1, 0) & 6 & 30 & 22.1 & 100.0 & 6.2 & 25.3 & 25.5 \\
        (0, 0, 1) & 20 & 5 & 7.2 & 0.0 & 8.7 & 19.4 & 28.1 \\
        (10, 1, 1) & 18 & 90 & 19.4 & 100.0 & 2.9 & 22.1 & 26.0 \\
        (11, 1, 1) & 18 & 90 & 19.4 & 100.0 & 2.9 & 22.1 & 26.0 \\
        (15, 1, 1) & 18 & 90 & 19.4 & 100.0 & 2.9 & 22.1 & 26.0 \\
        \bottomrule
    \end{tabular} 
    \caption{{
    Optimal CBCP and DBCP schemes for Pareto weights with the associated travel times (TTs) and fraction of users on the express lane.}}
    \label{table:ParetoResults}
\end{table}


\section{Conclusion and Future Work}
\label{sec: Conclusion and Future Work}

This work contrasts the effects of deploying CBCP and DBCP policies on traffic networks with tolled express lanes, when a subset of the users is eligible for travel assistance. 
\revision{Our theoretical analysis and empirical studies highlight the crucial role played by key user-dependent and network-specific parameters, such as users' VoTs and edge latency functions, in determining the efficacy of DBCP and CBCP policies in improving eligible users' express lane access 
}

As future work, we aim to extend our comparison study of CBCP and DBCP equilibria in Sec. \ref{sec: Budget vs Discount} to general networks, and identify conditions under which one type of policy outperforms the other in optimizing a pre-specified metric of societal welfare. We also aim to develop a principled search method for the optimal toll, budget, and discount values for CBCP and DBCP policies deployed on general traffic networks. Finally, we aim to design novel mechanisms, e.g., hybrid credit-discount based policies, that can outperform both purely CBCP and purely DBCP policies.

\bibliographystyle{abbrv}
\bibliography{refs}

\appendix

\section{Section \ref{sec: Equilibria Characterization} Proofs}
\label{sec: App A1, Section 4 Proofs}

\subsection{Proof of Theorem \ref{Thm: Convex Program, DBCP}}
\label{subsec: App A1, Proof of Thm Convex Program, DBCP}


For each $t \in [T]$, define the objective function $F^t: \Y^d \rightarrow \R$ by:
\begin{align*}
    F^t(y) &:= \sum_{e \in \edges} \Bigg[ \sum_{k=1}^2 \int_0^{x_{e,k,t}} \ell_{e,k}(w) \hspace{0.5mm} dw + \sum_{g \in \G_I} \frac{y_{e,1,t}^g \tau_{e,t}}{v_t^g} + \sum_{g \in \G_E} \frac{y_{e,1,t}^g (1 - \alpha_{e,t}) \tau_{e,t}}{v_t^g} \Bigg].
\end{align*}
Then the partial derivatives of $F$ with respect to $y$ are:
\begin{align*}
    \frac{\partial F^t}{\partial y_{e,1,t}^g} &= \ell_{e,1}(x_{e,1,t}) + \frac{\tau_{e,t}}{v_t^g}, \hspace{1cm} \forall \ y \in \G_I, e \in \edges, t \in [T], \\
    \frac{\partial F^t}{\partial y_{e,1,t}^g} &= \ell_{e,1}(x_{e,1,t}) + (1 - \alpha_{e,t}) \cdot \frac{\tau_{e,t}}{v_t^g}, \hspace{1cm} \forall \ y \in \G_E, e \in \edges, t \in [T], \\
    \frac{\partial F^t}{\partial y_{e,2,t}^g} &= \ell_{e,2}(x_{e,2,t}), \hspace{1cm} \forall \ g \in \G, e \in \edges, t \in [T].
\end{align*}
The first-order conditions for the minimizers of $F$ state that, for each $y \in \Y^d$:
\begin{align} \nonumber
    0 &\leq \sum_{e \in \edges} \Bigg[ \sum_{g \in \G_I} \Bigg( \ell_{e,1}(x_{e,1,t}^\star) + \frac{\tau_{e,t}}{v_t^g} \Bigg) (y_{e,1,t}^g - y_{e,1,t}^{\star g}) \\ \nonumber
    &\hspace{2cm} + \sum_{g \in \G_E} \Bigg( \ell_{e,1}(x_{e,1,t}^\star) + (1 - \alpha_{e,t}) \frac{\tau_{e,t}}{v_t^g} \Bigg) (y_{e,1,t}^g - y_{e,1,t}^{\star g}) \\ \nonumber
    &\hspace{2cm} + \sum_{g \in \G} \ell_{e,2}(x_{e,2,t}^\star) (y_{e,2,t}^g - y_{e,2,t}^{\star g}) \Bigg] \\ \label{Eqn: DBCP, First-Order Optimality Conditions}
    &= \sum_{g \in \G} \frac{1}{v_t^g} \sum_{e \in \edges} \sum_{k \in [2]} c_{e,k,t}^g(x_{e,k,t}^\star) \cdot (y_{e,k,t}^g - y_{e,k,t}^{\star g}).
\end{align}

Below, we show that \eqref{Eqn: DBCP, First-Order Optimality Conditions} is equivalent to the CBCP conditions.

~\\
\noindent
\say{Convex Program $\ \Rightarrow \ $ CBCP Equilibria}

To establish that \eqref{Eqn: DBCP, First-Order Optimality Conditions} yields the CBCP equilibrium conditions for eligible users, fix $t' \in [T]$, $g' \in \G$ arbitrarily, and let $y \in \Y^d$ be given such that $y_t^g = (y^\star)_t^g$ for each $g \in \G \backslash \{g'\}$. Then the above inequality becomes:
\begin{align*}
    0 &\leq \frac{1}{v_{t'}^{g'}} \sum_{e \in \edges} \sum_{k \in [2]} c_{e,k,t'}^{g'} (x_{e,k,t'}^\star) \cdot (y_{e,k,t'}^{g'} - y_{e,k,t'}^{\star g'}), \\
    \Ra \hspace{0.5mm} 0 &\leq \sum_{e \in \edges} \sum_{k \in [2]} c_{e,k,t'}^{g'} (x_{e,k,t'}^\star) \cdot (y_{e,k,t'}^{g'} - y_{e,k,t'}^{\star g'}),
\end{align*}
which are the CBCP equilibrium conditions for the discount setting.

~\\
\noindent
\say{CBCP Equilibria $\ \Rightarrow \ $ Convex Program}

Conversely, suppose $y^\star$ is a CBCP equilibrium, and for each $y \in \Y^d$, we have, at each time $t \in [T]$ and for each group $g \in \G$:
\begin{align*}
    &\sum_{e \in \edges} \sum_{k \in [2]} c_{e,k,t}^{g} (x_{e,k,t}^\star) \cdot (y_{e,k,t}^{g} - y_{e,k,t}^{\star g}) \geq 0.
\end{align*}
By multiplying $1/v_t^g$ on both sides and summing over all times $t \in [T]$ and all groups $g \in \G$, we recover \eqref{Eqn: DBCP, First-Order Optimality Conditions}.

\subsection{Proof of Theorem \ref{Thm: Convex Program, CBCP}}
\label{subsec: App A1, Proof of Thm Convex Program, CBCP}

Define the objective function $F: \Y^b \rightarrow \R$ by:
\begin{align*}
    F(y) &:= \sum_{t \in [T]} \sum_{e \in \edges} \Bigg[ \sum_{k=1}^2 \int_0^{x_{e,k,t}} \ell_{e,k}(w) \hspace{0.5mm} dw + \sum_{g \in \G} \frac{\hat y_{e,1,t}^g \tau_{e,t}}{v_t^g} \Bigg].
\end{align*}
Then the partial derivatives of $F$ with respect to $y$ are:
\begin{align*}
    \frac{\partial F}{\partial \hat y_{e,1,t}^g} &= \ell_{e,1}(x_{e,1,t}) + \frac{\tau_{e,t}}{v_t^g}, \hspace{1cm} \forall \ g \in \G, e \in \edges, t \in [T], \\
    \frac{\partial F}{\partial \tilde y_{e,1,t}^g} &= \ell_{e,1}(x_{e,1,t}), \hspace{1cm} \forall \ g \in \G, e \in \edges, t \in [T], \\
    \frac{\partial F}{\partial y_{e,2,t}^g} &= \ell_{e,2}(x_{e,2,t}), \hspace{1cm} \forall \ g \in \G, e \in \edges, t \in [T].
\end{align*}
From the first-order conditions for the minimizers of $F^b$:
\begin{align} \nonumber
    0 &\leq \sum_{t \in [T]} \sum_{e \in \edges} \Bigg[ \sum_{g \in \G} \Bigg( \ell_{e,1}(x_{e,1,t}^\star) + \frac{\tau_{e,t}}{v_t^g} \Bigg) (\hat y_{e,1,t}^g - \hat y_{e,1,t}^{\star g}) + \sum_{g \in \G} \ell_{e,1}(x_{e,1,t}^\star) \cdot (\tilde y_{e,1,t}^g - \tilde y_{e,1,t}^{\star g}) \\ \nonumber
    &\hspace{2cm} + \sum_{g \in \G} \ell_{e,2}(x_{e,2,t}^\star) (y_{e,2,t}^g - y_{e,2,t}^{\star g}) \Bigg] \\ \label{Eqn: CBCP, First-Order Optimality Conditions}
    &= \sum_{g \in \G_I} \sum_{t \in [T]} \frac{1}{v_t^g} \sum_{e \in \edges} \Big[ \hat c_{e,1,t}^g(x_{e,1,t}^\star) \cdot (\hat y_{e,1,t}^g - \hat y_{e,1,t}^{\star g}) + c_{e,2,t}^g(x_{e,2,t}^\star) \cdot (y_{e,2,t}^g - y_{e,2,t}^{\star g}) \Big] \\ \nonumber
    &\hspace{1cm} + \sum_{g \in \G_E} \frac{1}{v^g} \sum_{t \in [T]} \sum_{e \in \edges} \Big[ \hat c_{e,1,t}^g(x_{e,1,t}^\star) \cdot (\hat y_{e,1,t}^g - \hat y_{e,1,t}^{\star g}) + \tilde c_{e,1,t}^g(x_{e,1,t}^\star) \cdot (\tilde y_{e,1,t}^g - \tilde y_{e,1,t}^{\star g}) \\ \nonumber
    &\hspace{5cm} + c_{e,2,t}^g(x_{e,2,t}^\star) \cdot (y_{e,2,t}^g - y_{e,2,t}^{\star g}) \Big].
\end{align}

The rest of this proof shows that  \eqref{Eqn: DBCP, First-Order Optimality Conditions} is equivalent to the CBCP conditions.

~\\
\noindent
\say{Convex Program $\ \Rightarrow \ $ CBCP Equilibria}

To establish that \eqref{Eqn: CBCP, First-Order Optimality Conditions} yields the CBCP equilibrium conditions for ineligible users, fix $g' \in \G_I$ arbitrarily, and let $y \in \Y^b$ be given such that $y_t^g = (y^\star)_t^g$ for each $g \ne g'$ and $t \in [T]$. Then the above inequality becomes:
\begin{align*}
    0 &\leq \frac{1}{v_{t'}^{g'}} \sum_{e \in \edges} \Big[ \hat c_{e,1,t}^{g'} (x_{e,1,t}^\star) \cdot (\hat y_{e,1,t}^{g'} - \hat y_{e,1,t}^{\star g'}) + c_{e,2,t}^{g'} (x_{e,2,t}^\star) \cdot (y_{e,2,t}^{g'} - y_{e,2,t}^{\star g'}) \Big], \\
    \Ra \hspace{0.5mm} 0 &\leq \sum_{e \in \edges} \Big[ \hat c_{e,1,t}^{g'} (x_{e,1,t}^\star) \cdot (\hat y_{e,1,t}^{g'} - \hat y_{e,1,t}^{\star g'}) + c_{e,2,t}^{g'} (x_{e,2,t}^\star) \cdot (y_{e,2,t}^{g'} - y_{e,2,t}^{\star g'}) \Big],
\end{align*}
which are the CBCP equilibrium conditions for ineligible users for the budget setting.

To establish that \eqref{Eqn: CBCP, First-Order Optimality Conditions} yields the CBCP equilibrium conditions for eligible users, fix $g' \in \G_I$ arbitrarily, and let $y \in \Y^b$ be given such that $y_t^g = (y^\star)_t^g$ for each $g \ne g'$ and $t \in [T]$. Then the above inequality becomes:
\begin{align*}
    0 &\leq \frac{1}{v^{g'}} \sum_{t \in [T]} \sum_{e \in \edges} \Big[ \hat c_{e,1,t}^{g'} (x_{e,1,t}^\star) \cdot (\hat y_{e,1,t}^{g'} - \hat y_{e,1,t}^{\star g'}) + \tilde c_{e,1,t}^{g'} (x_{e,1,t}^\star) \cdot (\tilde y_{e,1,t}^{g'} - \tilde y_{e,1,t}^{\star g'}) \\
    &\hspace{4cm} + c_{e,2,t}^{g'} (x_{e,2,t}^\star) \cdot (y_{e,2,t}^{g'} - y_{e,2,t}^{\star g'}) \Big], \\
    \Ra \hspace{0.5mm} 0 &\leq \sum_{t \in [T]} \sum_{e \in \edges} \Big[ \hat c_{e,1,t}^{g'} (x_{e,1,t}^\star) \cdot (\hat y_{e,1,t}^{g'} - \hat y_{e,1,t}^{\star g'}) + \tilde c_{e,1,t}^{g'} (x_{e,1,t}^\star) \cdot (\tilde y_{e,1,t}^{g'} - \tilde y_{e,1,t}^{\star g'}) \\
    &\hspace{4cm} + c_{e,2,t}^{g'} (x_{e,2,t}^\star) \cdot (y_{e,2,t}^{g'} - y_{e,2,t}^{\star g'}) \Big].
\end{align*}
which are the CBCP equilibrium conditions for eligible users for the budget setting.

~\\
\noindent
\say{CBCP Equilibria $\ \Rightarrow \ $ Convex Program}

Conversely, suppose $y^\star$ is a CBCP equilibrium, and $y \in \Y$ satisfies, for each time $t \in [T]$ and each ineligible group $g \in \G_I$:
\begin{align*}
    &\sum_{e \in \edges} \Big[ \hat c_{e,1,t}^{g} (x_{e,1,t}^\star) \cdot (\hat y_{e,1,t}^{g} - \hat y_{e,1,t}^{\star g}) + c_{e,2,t}^{g} (x_{e,2,t}^\star) \cdot (y_{e,2,t}^{g} - y_{e,2,t}^{\star g}) \Big] \geq 0.
\end{align*}
and for each eligible group $g \in \G_E$:
\begin{align*}
    &\sum_{t \in [T]} \sum_{e \in \edges} \Big[ \hat c_{e,1,t}^{g} (x_{e,1,t}^\star) \cdot (\hat y_{e,1,t}^{g} - \hat y_{e,1,t}^{\star g}) + \tilde c_{e,1,t}^{g} (x_{e,1,t}^\star) \cdot (\tilde y_{e,1,t}^{g} - \tilde y_{e,1,t}^{\star g}) \\
    &\hspace{2cm} + c_{e,2,t}^{g} (x_{e,2,t}^\star) \cdot (y_{e,2,t}^{g} - y_{e,2,t}^{\star g}) \Big] \geq 0.
\end{align*}
By summing over all $t \in [T]$ and $g \in \G_I$ for the first inequality, and over all $g \in \G_E$ for the second inequality, we recover \eqref{Eqn: CBCP, First-Order Optimality Conditions}.

\subsection{Proof of Proposition \ref{Prop: Lane Flows Uniqueness, DBCP}}
\label{subsec: App A1, Proof of Prop Equilibrium Lane Flows Uniqueness, DBCP}

Let $y^\star, z^\star \in \Y$ be two DBCP-($\boldsymbol{\tau}, \boldsymbol{\alpha}$) equilibria, and let $x^\star(y^\star)$ and $x^\star(z^\star)$ denote the corresponding edge flows, respectively. By the convex program characterization for DBCP Equilibria (Theorem \ref{Thm: Convex Program, DBCP}), $x^\star(y^\star)$ and $x^\star(z^\star)$ must both satisfy \eqref{Eqn: DBCP, First-Order Optimality Conditions} for any $y \in \Y$. In particular:
\begin{align*}
    &\sum_{t=1}^T \sum_{e \in \edges} \Bigg[ \sum_{g \in \G_I} \Bigg(\ell_{e,1}(x_{e,1,t}^\star(y^\star)) + \frac{\tau_{e,t}}{v_t^g} \Bigg) \cdot (z_{e,1,t}^{g \star} - y_{e,1,t}^{g \star}) \\
    &\hspace{2.5cm} + \sum_{g \in \G_E} \Bigg(\ell_{e,1}(x_{e,1,t}^\star(y^\star)) + (1 - \alpha_{e,t}) \frac{\tau_{e,t}}{v_t^g} \Bigg) \cdot (z_{e,1,t}^{g \star} - y_{e,1,t}^{g \star}) \\
    &\hspace{2.5cm} + \sum_{g \in \G} \ell_{e,2}(x_{e,2,t}^\star (y^\star)) \cdot (z_{e,2,t}^{g \star} - y_{e,2,t}^{g \star}) \Bigg] \geq 0, \\
    &\sum_{t=1}^T \sum_{e \in \edges} \Bigg[ \sum_{g \in \G_I} \Bigg(\ell_{e,1}(x_{e,1,t}^\star(z^\star)) + \frac{\tau_{e,t}}{v_t^g} \Bigg) \cdot (y_{e,1,t}^{g \star} - z_{e,1,t}^{g \star}) \\
    &\hspace{2.5cm} + \sum_{g \in \G_E} \Bigg(\ell_{e,1}(x_{e,1,t}^\star(z^\star)) + (1 - \alpha_{e,t}) \frac{\tau_{e,t}}{v_t^g} \Bigg) \cdot (y_{e,1,t}^{g \star} - z_{e,1,t}^{g \star}) \\
    &\hspace{2.5cm} + \sum_{g \in \G} \ell_{e,2}(x_{e,2,t}^\star (z^\star)) \cdot (y_{e,2,t}^{g \star} - z_{e,2,t}^{g \star}) \Bigg] \geq 0.
\end{align*}
Summing the above two inequalities and rearranging terms gives:
\begin{align*}
    0 &\leq \sum_{t=1}^T \sum_{e \in \edges} \sum_{g \in \G} \Bigg[ \Big( \ell_{e,1}(x_{e,1,t}^\star(y^\star)) - \ell_{e,1}(x_{e,1,t}^\star(z^\star)) \Big) \cdot (z_{e,1,t}^{g \star} - y_{e,1,t}^{g \star}) \\
    &\hspace{3cm} + \Big( \ell_{e,2}(x_{e,2,t}^\star(y^\star)) - \ell_{e,2}(x_{e,2,t}^\star(z^\star)) \Big) \cdot (z_{e,2,t}^{g \star} - y_{e,2,t}^{g \star}) \Bigg] \\
    &= \sum_{t=1}^T \sum_{e \in \edges} \sum_{k=1}^2 \Bigg[ \Big( \ell_{e,k}(x_{e,k,t}^\star(y^\star)) - \ell_{e,k}(x_{e,k,t}^\star(z^\star)) \Big) \cdot \sum_{g \in \G} (z_{e,k,t}^{g \star} - y_{e,k,t}^{g \star}) \Bigg] \\
    &= \sum_{t=1}^T \sum_{e \in \edges} \sum_{k=1}^2 \Big( \ell_{e,k}(x_{e,k,t}^\star(y^\star)) - \ell_{e,k}(x_{e,k,t}^\star(z^\star)) \Big) \cdot \sum_{g \in \G} (x_{e,k,t}^{\star}(z^\star) - x_{e,k,t}^{\star}(y^\star)).
\end{align*}
Since $\ell_{e,k}$ is strictly increasing for each $e \in \edges$, $k \in [2]$, we conclude that $x_{e,k,t}^\star(y^\star) = x_{e,k,t}^\star(z^\star)$ for each $t \in [T]$ and $k \in [2]$. This concludes the proof.

\subsection{Proof of Proposition \ref{Prop: Lane Flows Uniqueness, CBCP}}
\label{subsec: App A1, Proof of Prop Equilibrium Lane Flows Uniqueness, CBCP}

\begin{proof}
Let $y^\star, z^\star \in \Y$ be two CBCP ($\boldsymbol{\tau}, B$) equilibria, and let $x^\star(y^\star)$ and $x^\star(z^\star)$ denote the corresponding edge flows, respectively. By the convex program characterization for CBCP Equilibria (Theorem \ref{Thm: Convex Program, CBCP}), $x^\star(y^\star)$ and $x^\star(z^\star)$ must both satisfy \eqref{Eqn: CBCP, First-Order Optimality Conditions} for any $y \in \Y$. In particular:
\begin{align*}
    &\sum_{t=1}^T \sum_{e \in \edges} \sum_{g \in \G} \Bigg[ \ell_{e,1}(x_{e,1,t}^\star(y^\star)) \cdot (\tilde z_{e,1,t}^{g \star} - \tilde y_{e,1,t}^{g \star}) + \Bigg(\ell_{e,1}(x_{e,1,t}^\star(y^\star)) + \frac{\tau_{e,t}}{v_t^g} \Bigg) \cdot (\hat z_{e,1,t}^{g \star} - \hat y_{e,1,t}^{g \star}) \\
    &\hspace{3cm} + \ell_{e,2}(x_{e,2,t}^\star (y^\star)) \cdot (z_{e,2,t}^{g \star} - y_{e,2,t}^{g \star}) \Bigg] \geq 0, \\
    &\sum_{t=1}^T \sum_{e \in \edges} \sum_{g \in \G} \Bigg[ \ell_{e,1}(x_{e,1,t}^\star(z^\star)) \cdot (\tilde y_{e,1,t}^{g \star} - \tilde z_{e,1,t}^{g \star}) + \Bigg(\ell_{e,1}(x_{e,1,t}^\star(z^\star)) + \frac{\tau_{e,t}}{v_t^g} \Bigg) \cdot (\hat y_{e,1,t}^{g \star} - \hat z_{e,1,t}^{g \star}) \\
    &\hspace{3cm} + \ell_{e,2}(x_{e,2,t}^\star (z^\star)) \cdot (y_{e,2,t}^{g \star} - z_{e,2,t}^{g \star}) \Bigg] \geq 0.
\end{align*}
Summing the above two inequalities and rearranging terms gives:
\begin{align*}
    0 &\leq \sum_{t=1}^T \sum_{e \in \edges} \sum_{g \in \G} \Bigg[ \Big( \ell_{e,1}(x_{e,1,t}^\star(y^\star)) - \ell_{e,1}(x_{e,1,t}^\star(z^\star)) \Big) \cdot \Big( (\tilde z_{e,1,t}^{g \star} - \tilde y_{e,1,t}^{g \star}) + (\hat z_{e,1,t}^{g \star} - \hat y_{e,1,t}^{g \star}) \Big) \\
    &\hspace{3cm} + \Big( \ell_{e,2}(x_{e,2,t}^\star(y^\star)) - \ell_{e,2}(x_{e,2,t}^\star(z^\star)) \Big) \cdot (z_{e,2,t}^{g \star} - y_{e,2,t}^{g \star}) \Bigg] \\
    &= \sum_{t=1}^T \sum_{e \in \edges} \Bigg[ \Big( \ell_{e,1}(x_{e,1,t}^\star(y^\star)) - \ell_{e,1}(x_{e,1,t}^\star(z^\star)) \Big) \cdot \sum_{g \in \G} \Big( (\tilde z_{e,1,t}^{g \star} - \tilde y_{e,1,t}^{g \star}) + (\hat z_{e,1,t}^{g \star} - \hat y_{e,1,t}^{g \star}) \Big) \\
    &\hspace{3cm} + \Big( \ell_{e,2}(x_{e,2,t}^\star(y^\star)) - \ell_{e,2}(x_{e,2,t}^\star(z^\star)) \Big) \cdot \sum_{g \in \G} (z_{e,2,t}^{g \star} - y_{e,2,t}^{g \star}) \Bigg] \\
    &= \sum_{t=1}^T \sum_{e \in \edges} \Bigg[ \Big( \ell_{e,1}(x_{e,1,t}^\star(y^\star)) - \ell_{e,1}(x_{e,1,t}^\star(z^\star)) \Big) \cdot \big(x_{e,1,t}^\star(z^\star) - x_{e,1,t}^\star(y^\star) \big) \\
    &\hspace{3cm} + \Big( \ell_{e,2}(x_{e,2,t}^\star(y^\star)) - \ell_{e,2}(x_{e,2,t}^\star(z^\star)) \Big) \cdot \big(x_{e,2,t}^\star(z^\star) - x_{e,2,t}^\star(y^\star) \big) \Bigg].
\end{align*}
Since $\ell_{e,1}$ and $\ell_{e,2}$ are strictly increasing, we conclude that $x_{e,1,t}^\star(y^\star) = x_{e,1,t}^\star(z^\star)$ and $x_{e,2,t}^\star(y^\star) = x_{e,2,t}^\star(z^\star)$ for each $t \in [T]$. This concludes the proof.
\end{proof}

\section{Section \ref{sec: Budget vs Discount} Proofs}
\label{sec: App A2, Section 5 Proofs}

\subsection{Proof of Theorem \ref{Thm: yc, yd, 10}}
\label{subsec: App A2, Proof of Thm yc, yd, 10}

\paragraph{Computing $y^C(\alpha)$:} We wish to compute $y^C(\alpha)$, to verify that \eqref{Eqn: yc, 10} is correct. First, to ensure that the expression given by \eqref{Eqn: yc, 10} is well-defined, we must first prove that there exists some unique $\alpha_1 \in (0, 1/2)$ such that:
\begin{align} \label{Eqn: alpha 1, Fixed point equation}
    \ell(\alpha_1) + \frac{\tau}{v^E} = \ell(1-\alpha_1).
\end{align}
To this end, let $f_1: [0, 1] \ra \R$ be given by:
\begin{align*}
    f_1(\alpha) &:= \ell(\alpha) + \frac{\tau}{v^E} - \ell(1-\alpha). 
\end{align*}
Since $\ell$ is strictly positive and strictly increasing, and $\tau < v^E(\ell(1) - \ell(0))$, we find that $f_1$ is also strictly increasing, from $f_1(0) < 0$ to $f_1(\frac{1}{2}) > 0$. Since $\ell$ is also differentiable, it is continuous, and thus so is $f_1$. Therefore, by the Intermediate Value Theorem, there exists a unique $\alpha_1 \in (0, \frac{1}{2})$ such that $f_1(\alpha_1) = 0$.

We now proceed to compute $y^C(\alpha)$. The Lagrangian $L: \R^3 \times \R^5 \ra \R$ corresponding to the convex program characterizing $(\tau, \alpha \tau)$-CBCP equilibria, i.e., \eqref{Eqn: Convex Program Statement, CBCP}, is given by:
\begin{align*}
    L(\tilde y_1, \hat y_1, y_2, \mu, \lambda, \tilde s_1, \hat s_1, s_2) &:= \int_0^{\tilde y_1 + \hat y_1} \ell(w) \ dw + \int_0^{y_2} \ell(w) \ dw + \frac{\hat y_1 \tau}{v^E} + \mu(\tilde y_1 - \alpha) \\
    &\hspace{1cm} - \lambda(\tilde y_1 + \hat y_1 + y_2 - 1) - \tilde s_1 \tilde y_1 - \hat s_1 \hat y_1 - s_2 y_2. 
\end{align*}
The corresponding KKT conditions are:
\begin{alignat}{2} \nonumber
    0 &= \frac{\partial L}{\partial \tilde y_1} = \ell(\tilde y_1 + \hat y_1) + \mu - \lambda - \tilde s_1, \hspace{1cm} && \tilde s_1 \geq 0, \hspace{5mm} \tilde s_1 \tilde y_1 = 0, \\ \nonumber
    0 &= \frac{\partial L}{\partial \hat y_1} = \ell(\tilde y_1 + \hat y_1) + \frac{\tau}{v^E} - \lambda - \hat s_1, \hspace{5mm} && \hat s_1 \geq 0, \hspace{5mm} \hat s_1 \hat y_1 = 0, \\ \nonumber
    0 &= \frac{\partial L}{\partial y_2} = \ell(y_2) - \lambda - s_2, &&s_2 \geq 0, \hspace{5mm} s_2 y_2 = 0, \\ \nonumber
    \mu &\geq 0, \hspace{5mm} \mu(\tilde y_1 - \alpha) = 0, \hspace{5mm} \tilde y_1^E \leq \alpha.
\end{alignat}
Rearranging, we obtain:
\begin{alignat}{2} \label{Eqn: KKT, 10, CBCP, 1}
    \ell(\tilde y_1 + \hat y_1) + \mu - \lambda &= \tilde s_1 \geq 0, \hspace{1cm} &&\tilde s_1 \tilde y_1 = 0, \\ \label{Eqn: KKT, 10, CBCP, 2}
    \ell(\tilde y_1 + \hat y_1) + \frac{\tau}{v^E} - \lambda &= \hat s_1 \geq 0, &&\hat s_1 \hat y_1 = 0, \\ \label{Eqn: KKT, 10, CBCP, 3}
    \ell(y_2) - \lambda &= s_2 \geq 0, &&s_2 y_2 = 0, \\ \label{Eqn: KKT, 10, CBCP, 4}
    \mu(\tilde y_1 - \alpha) &= 0, &&\mu \geq 0.
\end{alignat}
When $\alpha \in (0, \alpha_1)$, the tuple:
\begin{align*}
    (\tilde y_1, \hat y_1, y_2, \mu, 
    \lambda, \tilde s_1, \hat s_1, s_2) &= \left( \alpha, \alpha_1 - \alpha, 1 - \alpha, \frac{\tau}{v^E}, \ell(\alpha_1) + \frac{\tau}{v^E}, 0, 0, 0 \right)
\end{align*}
satisfies \eqref{Eqn: KKT, 10, CBCP, 1}-\eqref{Eqn: KKT, 10, CBCP, 4}. Similarly, when $\alpha \in (\alpha_1, 1/2)$, the tuple:
\begin{align*}
    (\tilde y_1, \hat y_1, y_2, \mu, 
    \lambda, \tilde s_1, \hat s_1, s_2) &= \left( \alpha, 0, 1 - \alpha, \ell(1-\alpha) - \ell(\alpha), \ell(1-\alpha), 0, \ell(\alpha) + \frac{\tau}{v^E} - \ell(1-\alpha), 0 \right)
\end{align*}
satisfies \eqref{Eqn: KKT, 10, CBCP, 1}-\eqref{Eqn: KKT, 10, CBCP, 4}. Finally, when $\alpha \in (1/2, 1)$, the tuple:
\begin{align*}
    (\tilde y_1, \hat y_1, y_2, \mu, 
    \lambda, \tilde s_1, \hat s_1, s_2) &= \left( \frac{1}{2}, 0, \frac{1}{2}, 0, \ell\left( \frac{1}{2} \right), 0, \frac{\tau}{v^E}, 0 \right)
\end{align*}
satisfies \eqref{Eqn: KKT, 10, CBCP, 1}-\eqref{Eqn: KKT, 10, CBCP, 4}.  Since the entire user population consists of a single group of eligible users, the equilibrium flows of eligible users on the express lane equal the aggregate express lane equilibrium flows, which are unique (Proposition \ref{Prop: Lane Flows Uniqueness, CBCP}). This implies that:
\begin{align*} 
    y^C(\alpha) &= \tilde y_1 + \hat y_1 = \begin{cases}
        \alpha_1, \hspace{5mm} &\alpha \in (0, \alpha_1), \\
        \alpha, \hspace{5mm} &\alpha \in (0, 1/2), \\
        1/2, \hspace{5mm} &\alpha \in (1/2, 0),
    \end{cases}
\end{align*}
as claimed.



\paragraph{Computing $y^D(\alpha)$:}
We first establish that, for each $\alpha \in [0, 1]$, there exists a unique $y^\star(\alpha) \in (0, 1/2)$ such that:
\begin{align} \label{Eqn: yd alpha, 10, Fixed point equation}
\ell(y^\star(\alpha)) + \frac{(1-\alpha)\tau}{v^E} = \ell(1-y^\star(\alpha)).
\end{align}
Then, we prove that $y^\star(\alpha)$ is continuously differentiable, strictly increasing, and strictly convex. Finally, we show that $y^D(\alpha) = y^\star(\alpha)$.

First, fix $\alpha \in [0, 1]$ arbitrarily, and define $F_1: \R \times [0, 1/2] \ra \R$ by:
\begin{align*}
    F_1(y, \alpha) &:= \ell(y) + \frac{(1-\alpha) \tau}{v^E} - \ell(1-y).
\end{align*}
Again, since $\ell$ is strictly positive and strictly increasing, and $\tau < v^E(\ell(1) - \ell(0))$, we find that $F_1(\cdot, \alpha)$ is also strictly increasing, from $F_1(0, \alpha) < 0$ to $F_1(\frac{1}{2}, \alpha) \geq 0$. Since $\ell$ is also differentiable, it is continuous, and thus so is $F_1(\cdot, \alpha)$. Therefore, by the Intermediate Value Theorem, there exists a unique $y \in (0, \frac{1}{2})$ such that $F_1(y, \alpha) = 0$. 

To show that $y^\star(\alpha)$ is strictly increasing and convex, we apply the Implicit Function Theorem to $F_1$. More precisely, note that for each $y \in \R, \alpha \in [0, 1/2]$:
\begin{align*}
    \frac{\partial F_1}{\partial y}(y, \alpha) &= \ell'(y) + \ell'(1-y) > 0, \\
    \frac{\partial F_1}{\partial \alpha}(y, \alpha) &= -\frac{\tau}{v^E}.
\end{align*}
Thus, by the Implicit Function Theorem, $y^\star$ is continuously differentiable in $\alpha$, and:
\begin{align*}
    \frac{dy^\star}{d\alpha}(\alpha) &= \left[ \frac{\partial F_1}{\partial y}(y^\star(\alpha), \alpha) \right]^{-1} \frac{\partial F_1}{\partial \alpha}(y^\star(\alpha), \alpha) \\
    &= \frac{\tau}{v^E \big[\ell'(1-y^\star(\alpha)) + \ell'(y^\star(\alpha)) \big]}, \\
    \frac{d^2 y^\star}{d\alpha^2}(\alpha) &= \frac{-\tau}{v^E\big[ \ell'(1-y^\star(\alpha)) + \ell'(y^\star(\alpha)) \big]^2} \cdot \Bigg[ - \frac{d^2 \ell}{dy^2} (1 - y^\star(\alpha)) + \frac{d^2 \ell}{dy^2}(y^\star(\alpha)) \Bigg] \frac{dy^\star}{d \alpha}(\alpha).
\end{align*}
Since the first, second, and third derivatives of the latency function $\ell(\cdot)$ are all strictly positive, we have $\frac{dy^\star}{d \alpha}(\alpha) > 0$ and $\frac{d^2 y^\star}{d\alpha^2}(\alpha) > 0$ for each $\alpha \in (0, 1/2)$, so $y^\star$ is a strictly increasing and strictly convex function.

Next, we show that $y^D(\alpha) = y^\star(\alpha)$. The Lagrangian corresponding to the convex program characterizing $(\tau, \alpha)$-DBCP equilibria, i.e., \eqref{Eqn: Convex Program Statement, DBCP}, is given by:
\begin{align*}
    L(y_1, y_2, \lambda, s_1, s_2) &:= \int_0^{y_1} \ell(w) \ dw + \int_0^{y_2} \ell(w) \ dw + \frac{y_1 (1-\alpha) \tau}{v^E} \\
    &\hspace{1cm} - \lambda(y_1 + y_2 - 1) - s_1 y_1 - s_2 y_2. 
\end{align*}
The corresponding KKT conditions are:
\begin{alignat}{2} \nonumber
    0 &= \frac{\partial L}{\partial y_1} = \ell(y_1) + \frac{(1-\alpha) \tau}{v^E} - \lambda - s_1, \hspace{5mm} &&s_1 \geq 0, \hspace{5mm} s_1 y_1 = 0, \\ \nonumber
    0 &= \frac{\partial L}{\partial y_2} = \ell(y_2) - \lambda - s_2, &&s_2 \geq 0, \hspace{5mm} s_2 y_2 = 0.
\end{alignat}
Rearranging, we obtain:
\begin{alignat}{2} \label{Eqn: KKT, 10, DBCP, 1}
    \ell(y_1) + \frac{(1-\alpha)\tau}{v^E} - \lambda &= s_1 \geq 0, \hspace{5mm} &&s_1 y_1 = 0, \\ \label{Eqn: KKT, 10, DBCP, 2}
    \ell(y_2) - \lambda &= s_2 \geq 0, &&s_2 y_2 = 0, \\ \label{Eqn: KKT, 10, DBCP, 3}
    \mu(\tilde y_1 - \alpha) &= 0, &&\mu \geq 0.
\end{alignat}
For each $\alpha \in [0, 1]$, the tuple:
\begin{align*}
    (y_1, y_2, \lambda, s_1, s_2) &= \left( y^\star(\alpha), 1 - y^\star(\alpha), \ell\big(1 - y^\star(\alpha) \big), 0, 0 \right)
\end{align*}
satisfies \eqref{Eqn: KKT, 10, CBCP, 1}-\eqref{Eqn: KKT, 10, CBCP, 4}. Again, since the entire user population consists of a single group of eligible users, the equilibrium flows of eligible users on the express lane equal the aggregate express lane equilibrium flows, which are unique (Proposition \ref{Prop: Lane Flows Uniqueness, CBCP}). This implies that $y^D(\alpha) = y^\star(\alpha)$, as claimed.

\subsection{Proof of Theorem \ref{Thm: yc vs. yd, 10}}
\label{subsec: App A2, Proof of Thm yc vs. yd, 10}

Since $y^C(0) = y^D(0) = \alpha_1$, $y^C(1) = y^D(1) = 1/2$, and $y^D(\cdot)$ is strictly increasing, for each $\alpha \in (0, \alpha_1)$, we have $y^C(\alpha) = \alpha_1 < y^D(\alpha)$, and for each $\alpha \in (1/2, 1)$, we have $y^C(\alpha) = 1/2 > y^D(\alpha)$. In particular, $y^C(\alpha_1) - y^D(\alpha_1) < 0$ and $y^C(1/2) - y^D(1/2) > 0$. Applying the Intermediate Value Theorem to the continuous map $\alpha \mapsto y^C(\alpha) - y^D(\alpha)$, we find that there exists some $\alpha_2 \in (\alpha_1, 1/2)$ such that $y^C(\alpha_2) = y^D(\alpha_2)$. We next show that $\alpha_2$ is unique. Suppose by contradiction that there exists some $\bar \alpha_2 \ne \alpha_2$ such that $y^C(\bar \alpha_2) = y^D(\bar \alpha_2)$. We will assume, without loss of generality, that $\bar \alpha_2 > \alpha_2$. Then, since $y^D$ is convex, we have:
\begin{align*}
    y^D(\bar \alpha_2) &\leq \frac{1/2 - \bar \alpha_2}{1/2 - \alpha_2} y^D(\alpha_2) + \frac{\bar \alpha_2 - \alpha_2}{1/2 - \alpha_2} y^D(1/2).
\end{align*}
Substituting $y^D(\bar \alpha_2) = y^C(\bar \alpha_2) = \bar \alpha_2$, $y^D(\alpha_2) = y^C(\alpha_2) = \alpha_2$ and rearranging terms, we find that $y^D(1/2) \geq 1/2 = y^D(1)$, a contradiction to the fact that $y^D$ is strictly increasing. Thus, $\alpha_2$ is unique.

\subsection{Proof of Theorem \ref{Thm: yc, yd, 11}}
\label{subsec: App A2, Proof of Thm yc, yd, 11}

\paragraph{Computing $y^C(\alpha)$:}
To compute $y^C(\alpha)$, we first write the Lagrangian corresponding to the convex program characterizing $(\tau, \alpha)$-DBCP equilibria, i.e., \eqref{Eqn: Convex Program Statement, CBCP}, given by:
\begin{align*}
    &L(\tilde y_1^E, \hat y_1^E, \hat y_1^I, y_2^E, y_2^I, \mu, \lambda^E, \lambda^I, \tilde s_1^E, \hat s_1^E, s_2^E, \hat s_1^I, s_2^I) \\
    := \ &\int_0^{\tilde y_1 + \hat y_1 + \hat y_1^I} \ell(w) \ dw + \int_0^{y_2^E + y_2^I} \ell(w) \ dw + \frac{\hat y_1^E \tau}{v^E} + \frac{\hat y_1^I \tau}{v^I} \\
    &\hspace{1cm} + \mu(\tilde y_1 - \alpha) - \lambda^E(\tilde y_1^E + \hat y_1^E + y_2^E - 1) - \lambda^I(\hat y_1^I + y_2^I - 1) \\
    &\hspace{1cm} - \tilde s_1^E \tilde y_1^E - \hat s_1^E \hat y_1^E - s_2^E y_2^E - \hat s_1^I \hat y_1^I - s_2^I y_2^I. 
\end{align*}
The corresponding KKT conditions are:
\begin{alignat}{2} \nonumber
    0 &= \frac{\partial L}{\partial \tilde y_1^E} = \ell(\tilde y_1^E + \hat y_1^E + \hat y_1^I) + \mu - \lambda^E - \tilde s_1^E, \hspace{1cm} && \tilde s_1^E \geq 0, \hspace{5mm} \tilde s_1^E \tilde y_1^E = 0, \\ \nonumber
    0 &= \frac{\partial L}{\partial \hat y_1^E} = \ell(\tilde y_1^E + \hat y_1^E + \hat y_1^I) + \frac{\tau}{v^E} - \lambda^E - \hat s_1^E, \hspace{5mm} && \hat s_1^E \geq 0, \hspace{5mm} \hat s_1^E \hat y_1^E = 0, \\ \nonumber
    0 &= \frac{\partial L}{\partial y_2^E} = \ell(y_2^E + y_2^I) - \lambda^E - s_2^E, &&s_2^E \geq 0, \hspace{5mm} s_2^E y_2^E = 0, \\ \nonumber
    0 &= \frac{\partial L}{\partial \hat y_1^I} = \ell(\tilde y_1^E + \hat y_1^E + \hat y_1^I) + \frac{\tau}{v^I} - \lambda^I - \hat s_1^I, \hspace{5mm} && \hat s_1^I \geq 0, \hspace{5mm} \hat s_1^I \hat y_1^I = 0, \\ \nonumber
    0 &= \frac{\partial L}{\partial y_2^I} = \ell(y_2^E + y_2^I) - \lambda^I - s_2^I, &&s_2^I \geq 0, \hspace{5mm} s_2^I y_2^I = 0, \\ \nonumber
    \mu &\geq 0, \hspace{5mm} \mu(\tilde y_1^E - \alpha) = 0, \hspace{5mm} \tilde y_1^E \leq \alpha.
\end{alignat}
Rearranging, we obtain:
\begin{alignat}{2} \label{Eqn: KKT, 11, CBCP, 1}
    \ell(\tilde y_1^E + \hat y_1^E + \hat y_1^I) + \mu - \lambda^E &= \tilde s_1^E \geq 0, \hspace{1cm} &&\tilde s_1^E \tilde y_1^E = 0, \\ \label{Eqn: KKT, 11, CBCP, 2}
    \ell(\tilde y_1^E + \hat y_1^E + \hat y_1^I) + \frac{\tau}{v^E} - \lambda^E &= \hat s_1^E \geq 0, &&\hat s_1^E \hat y_1^E = 0, \\ \label{Eqn: KKT, 11, CBCP, 3}
    \ell(y_2^E + y_2^I) - \lambda^E &= s_2^E \geq 0, &&s_2^E y_2^E = 0, \\ \label{Eqn: KKT, 11, CBCP, 4}
    \ell(\tilde y_1^E + \hat y_1^E + \hat y_1^I) + \frac{\tau}{v^I} - \lambda^I &= \hat s_1^I \geq 0, &&\hat s_1^I \hat y_1^I = 0, \\ \label{Eqn: KKT, 11, CBCP, 5}
    \ell(y_2^E + y_2^I) - \lambda^I &= s_2^I \geq 0, &&s_2^I y_2^I = 0, \\ \label{Eqn: KKT, 11, CBCP, 6}
    \mu(\tilde y_1^E - \alpha) &= 0, &&\mu \geq 0, \tilde y_1^E \leq \alpha.
\end{alignat}
From \eqref{Eqn: KKT, 11, CBCP, 1}-\eqref{Eqn: KKT, 11, CBCP, 6}, we find that:
\begin{align*}
    \lambda^E &= \min\left\{ \ell(\tilde y_1^E + \hat y_1^E + \hat y_1^I) + \mu, \ell(\tilde y_1^E + \hat y_1^E + \hat y_1^I) + \frac{\tau}{v^E}, \ell(y_2^E + y_2^I) \right\}, \\
    \lambda^I &= \min\left\{ \ell(\tilde y_1^E + \hat y_1^E + \hat y_1^I) + \frac{\tau}{v^I}, \ell(y_2^E + y_2^I)\right\}.
\end{align*}
Note that if $y^C(\alpha)$ is well-defined, it equals $\tilde y_1 + \hat y_1$ at the given value of $\alpha$. Below, we show that $y^C(\alpha) = \alpha$ by proving that $\tilde y_1 = \alpha$ and $\hat y_1 = 0$ at each value of $\alpha$.

If $\tilde y_1^E < \alpha$, then $\mu = 0$, so:
\begin{align*}
    \hat s_1^E = \ell(\tilde y_1^E + \hat y_1^E + \hat y_1^I) + \frac{\tau}{v^E} - \lambda^E > \ell(\tilde y_1^E + \hat y_1^E + \hat y_1^I) - \lambda^E = \tilde s_1^E \geq 0,
\end{align*}
and thus $\hat y_1^E = 0$. As a result, $y_2^E > 1 - \tilde y_1^E - \hat y_1^E > 1 - \alpha > 0$ and $s_2^E = 0$, so:
\begin{align*}
    \ell(\tilde y_1^E + \hat y_1^E + \hat y_1^I) \geq \lambda^E = \ell(y_2^E + y_2^I).
\end{align*}
Then, we have:
\begin{align*}
    \hat s_1^I = \ell(\tilde y_1^E + \hat y_1^E + \hat y_1^I) + \frac{\tau}{v^E} - \lambda^E > \ell(y_2^E + y_2^I) - \lambda^E = s_2^E \geq 0,
\end{align*}
Thus, $\hat y_1^I = 0$, so $y_2^I = 1$ and $s_2^I = 0$. In this case, $y_2^E = 1 - \tilde y_1^E - \hat y_1^E = 1 - \tilde y_1^E$, and so \eqref{Eqn: KKT, 11, CBCP, 1} and \eqref{Eqn: KKT, 11, CBCP, 3} become:
\begin{align*}
    \ell(\tilde y_1^E) - \lambda^E = \tilde s_1^E &\geq 0, \\
    \ell(2 - \tilde y_1^E) - \lambda^E = s_2^E &= 0.
\end{align*}
Rearranging terms, we find that $\ell(\tilde y_1^E) \geq \lambda^E = \ell(2 - \tilde y_1^E)$, a contradiction, since $\tilde y_1^E < \alpha \leq 1$, and $\ell$ is strictly increasing. We conclude that $\tilde y_1^E = \alpha$.

Next, suppose by contradiction that $\hat y_1^E > 0$. Then $\hat s_1^E = 0$, and:
\begin{align*}
    &\ell(\tilde y_1^E + \hat y_1^E + \hat y_1^I) + \frac{\tau}{v^E} = \lambda^E \leq \ell(y_2^E + y_2^I).
\end{align*}
Since $v^E < v^I$, we have:
\begin{align} \label{Eqn: Contra, 1}
    &\ell(\tilde y_1^E + \hat y_1^E + \hat y_1^I) + \frac{\tau}{v^I} < \ell(\tilde y_1^E + \hat y_1^E + \hat y_1^I) + \frac{\tau}{v^E} \leq \ell(y_2^E + y_2^I), \\ \nonumber
    \Ra \ &s_2^I = \ell(y_2^E + y_2^I) - \lambda^I > \ell(\tilde y_1^E + \hat y_1^E + \hat y_1^I) + \frac{\tau}{v^I} - \lambda^I = \hat s_1^I \geq 0.
\end{align}
so $y_2^I = 0$ and $\hat y_1^I = 1$. But then:
\begin{align*}
    &\ell(\tilde y_1^E + \hat y_1^E + \hat y_1^I) + \frac{\tau}{v^I} > \ell(1) \geq \ell(y_2^E) = \ell(y_2^E + y_2^I),
\end{align*}
a contradiction to \eqref{Eqn: Contra, 1}. We conclude that $\hat y_1^E = 0$. As a result, $y^C(\alpha) = \tilde y_1^E + \hat y_1^E = \alpha + 0 = \alpha$.

\paragraph{Computing $y^D(\alpha)$:}
We first establish that, for each $\alpha \in [0, 1]$, there exists a unique $y^\dagger(\alpha) \in (0, 1/2)$ such that:
\begin{align} \label{Eqn: yd alpha, 11, Fixed point equation}
    \ell(y^\dagger(\alpha)) + \frac{(1-\alpha)\tau}{v^E} = \ell(2 - y^\dagger(\alpha)).
\end{align}
Then, we prove that $y^\dagger(\alpha)$ is continuously differentiable, strictly increasing, and strictly convex. Finally, we show that $y^D(\alpha) = y^\dagger(\alpha)$.

First, fix $\alpha \in [0, 1]$ arbitrarily, and define $F_2: \R \times [0, 1/2] \ra \R$ by:
\begin{align*}
    F_2(y, \alpha) &:= \ell(y) + \frac{(1-\alpha) \tau}{v^E} - \ell(2 - y).
\end{align*}
Since $\ell$ is strictly positive and strictly increasing, and $\tau < v^E(\ell(2) - \ell(0))$, we find that $F_2(\cdot, \alpha)$ is also strictly increasing, from $F_2(0, \alpha) < 0$ to $F_2(1, \alpha) \geq 0$. Since $\ell$ is also differentiable, it is continuous, and thus so is $F_2(\cdot, \alpha)$. Therefore, by the Intermediate Value Theorem, there exists a unique $y^\dagger(\alpha) \in (0, 1]$ such that $F_2(y^\dagger(\alpha), \alpha) = 0$.

To show that $y^\dagger(\alpha)$ is strictly increasing and convex, we apply the Implicit Function Theorem to $F$. More precisely, note that for each $y \in \R, \alpha \in [0, 1/2]$:
\begin{align*}
    \frac{\partial F_2}{\partial y}(y, \alpha) &= \ell'(y) + \ell'(2-y) > 0, \\
    \frac{\partial F_2}{\partial \alpha}(y, \alpha) &= -\frac{\tau}{v^E}.
\end{align*}
Thus, by the Implicit Function Theorem, $y^\dagger$ is continuously differentiable in $\alpha$, and:
\begin{align} \nonumber
    \frac{dy^\dagger}{d\alpha}(\alpha) &= \left[ \frac{\partial F_2}{\partial y}(y^\dagger(\alpha), \alpha) \right]^{-1} \frac{\partial F_2}{\partial \alpha}(y^\dagger(\alpha), \alpha) \\ \label{Eqn: y dagger, first derivative}
    &= \frac{\tau}{v^E \big[\ell'(2 -y^\dagger(\alpha)) + \ell'(y^\dagger(\alpha)) \big]}, \\ \label{Eqn: y dagger, second derivative}
    \frac{d^2 y^\dagger}{d\alpha^2}(\alpha) &= \frac{-\tau}{v^E \big[ \ell'(2 - y^\dagger(\alpha)) + \ell'(y^\dagger(\alpha)) \big]^2} \cdot \Bigg[ - \frac{d^2 \ell}{dy^2} (2 - y^\dagger(\alpha)) \frac{dy^\dagger}{d \alpha} + \frac{d^2 \ell}{dy^2}(y^\dagger(\alpha)) \frac{dy^\dagger}{d \alpha} \Bigg]
\end{align}
Since the first, second, and third derivatives of the latency function $\ell(\cdot)$ are all strictly positive, we have $\frac{dy^\dagger}{d \alpha}(\alpha) > 0$ and $\frac{d^2 y^\dagger}{d\alpha^2}(\alpha) > 0$ for each $\alpha \in (0, 1/2)$, so $y^\dagger$ is a strictly increasing and strictly convex function.

Below, we prove that $y^D(\alpha) = y^\dagger(\alpha)$. To this end, note that the Lagrangian corresponding to the convex program characterizing $(\tau, \alpha)$-DBCP equilibria, i.e., \eqref{Eqn: Convex Program Statement, DBCP}, is given by:
\begin{align*}
    &L(y_1^E, y_1^I, y_2^E, y_2^I, \mu, \lambda^E, \lambda^I, s_1^E, s_1^I, s_2^E, s_2^I) \\
    := \ &\int_0^{y_1^E + y_1^I} \ell(w) \ dw + \int_0^{y_2^E + y_2^I} \ell(w) \ dw + \frac{y_1^E (1-\alpha) \tau}{v^E} + \frac{y_1^I \tau}{v^I} \\
    &\hspace{1cm} - \lambda^E(y_1^E + y_2^E - 1) - \lambda^I(y_1^I + y_2^I - 1) \\
    &\hspace{1cm} - s_1^E y_1^E - s_2^E y_2^E - s_1^I y_1^I - s_2^I y_2^I. 
\end{align*}
The corresponding KKT conditions are:
\begin{alignat}{2} \nonumber
    0 &= \frac{\partial L}{\partial y_1^E} = \ell(y_1^E + y_1^I) + \frac{(1-\alpha) \tau}{v^E} - \lambda^E - s_1^E, \hspace{5mm} &&s_1^E \geq 0, \hspace{5mm} s_1^E y_1^E = 0, \\ \nonumber
    0 &= \frac{\partial L}{\partial y_2^E} = \ell(y_2^E + y_2^I) - \lambda^E - s_2^E, &&s_2^E \geq 0, \hspace{5mm} s_2^E y_2^E = 0, \\
    0 &= \frac{\partial L}{\partial y_1^I} = \ell(y_1^E + y_1^I) + \frac{\tau}{v^I} - \lambda^I - s_1^I, \hspace{5mm} &&s_1^I \geq 0, \hspace{5mm} s_1^I y_1^I = 0, \\ \nonumber
    0 &= \frac{\partial L}{\partial y_2^I} = \ell(y_2^E + y_2^I) - \lambda^I - s_2^I, &&s_2^I \geq 0, \hspace{5mm} s_2^I y_2^I = 0.
\end{alignat}
Rearranging, we obtain:
\begin{alignat}{2} \label{Eqn: KKT, 11, DBCP, 1}
    \ell(y_1^E + y_1^I) + \frac{(1-\alpha)\tau}{v^E} - \lambda^E &= s_1^E \geq 0, \hspace{5mm} &&s_1^E y_1^E = 0, \\ \label{Eqn: KKT, 11, DBCP, 2}
    \ell(y_2^E + y_2^I) - \lambda^E &= s_2^E \geq 0, &&s_2^E y_2^E = 0, \\ \label{Eqn: KKT, 11, DBCP, 3}
    \ell(y_1^E + y_1^I) + \frac{(1-\alpha)\tau}{v^I} - \lambda^I &= s_1^I \geq 0, \hspace{5mm} &&s_1^I y_1^I = 0, \\ \label{Eqn: KKT, 11, DBCP, 4}
    \ell(y_2^E + y_2^I) - \lambda^I &= s_2^I \geq 0, &&s_2^I y_2^I = 0, \\ \label{Eqn: KKT, 11, DBCP, 5}
    \mu(\tilde y_1 - \alpha) &= 0, &&\mu \geq 0.
\end{alignat}
We separate the analysis of \eqref{Eqn: KKT, 11, DBCP, 1}-\eqref{Eqn: KKT, 11, DBCP, 4} into two cases below---The $\alpha < 1 - v^E/ v^I$ case and the $\alpha > 1 - v^E/ v^I$ case. 

First, suppose $\alpha < 1 - v^E/ v^I$, and suppose by contradiction that $y_1^E > 0$. Then $s_1^E = 0$, so:
\begin{align*}
    &\ell(y_1^E + y_1^I) + \frac{(1-\alpha)\tau}{v^E} = \lambda^E \leq \ell(y_2^E + y_2^I), \\
    \Ra \ &\ell(y_1^E + y_1^I) + \frac{\tau}{v^I} < \ell(y_1^E + y_1^I) + \frac{(1-\alpha)\tau}{v^E} \leq \ell(y_2^E + y_2^I), \\
    \Ra \ &s_2^I = \ell(y_2^E + y_2^I) - \lambda^I > \ell(y_1^E + y_1^I) + \frac{\tau}{v^I} - \lambda^I = s^I \geq 0,
\end{align*}
so $y_2^I = 0$ and $y_1^I = 1$. But then:
\begin{align*}
    &\ell(y_1^E + y_1^I) + \frac{\tau}{v^I} > \ell(1) > \ell(y_2^E) = \ell(y_2^E + y_2^I),
\end{align*}
a contradiction. Thus, $y^D(\alpha) = y_1^E = 0$.

Next, suppose $\alpha > 1 - v^E/v^I$, and suppose by contradiction that $y_1^I > 0$. Then $s_1^I = 0$, so:
\begin{align} \label{Eqn: Contra, 2}
    &\ell(y_1^E + y_1^I) + \frac{\tau}{v^I} = s_1^I + \lambda^I = \lambda^I \leq \ell(y_2^E + y_2^I), \\ \nonumber
    \Ra \ &\ell(y_1^E + y_1^I) + \frac{(1-\alpha)\tau}{v^E} < \ell(y_1^E + y_1^I) + \frac{\tau}{v^I} \leq \ell(y_2^E + y_2^I), \\
    \Ra \ &s_2^E = \ell(y_2^E + y_2^I) - \lambda^E > \ell(y_1^E + y_1^I) + \frac{\tau}{v^E} - \lambda^E = s_1^E \geq 0, 
\end{align}
so $y_2^E = 0$ and $y_1^E = 1$, and thus:
\begin{align*}
    &\ell(y_1^E + y_1^I) + \frac{(1-\alpha) \tau}{v^I} > \ell(1) > \ell(y_2^I) = \ell(y_2^E + y_2^I),
\end{align*}
a contradiction to \eqref{Eqn: Contra, 2}. So $y^D(\alpha) = y_1^E$, and from \eqref{Eqn: KKT, 11, DBCP, 1} and \eqref{Eqn: KKT, 11, DBCP, 2} become:
\begin{align} \label{Eqn: KKT simplified, 11, DBCP, 1}
    \ell(y_1^E) + \frac{(1-\alpha) \tau}{v^E} - \lambda^E &= s_1^E \geq 0, \hspace{5mm} s_1^E y_1^E = 0, \\ \label{Eqn: KKT simplified, 11, DBCP, 2}
    \ell(2 - y_1^E) - \lambda^E &= s_2^E \geq 0, \hspace{5mm} s_2^E y_2^E = 0,
\end{align}
Now, if $y_1^E = 1$, then $s_1^E = 0$, and as a result, \eqref{Eqn: KKT simplified, 11, DBCP, 1} and \eqref{Eqn: KKT simplified, 11, DBCP, 2} imply:
\begin{align*}
    \ell(1) + \frac{(1-\alpha)\tau}{v^E} = \lambda^E \leq \ell(1),
\end{align*}
a contradiction. Thus, $y_1^E < 1$. Similarly, if $y_2^E = 1$, then $s_2^E = 0$ and $y_1^E = 0$, and as a result, \eqref{Eqn: KKT simplified, 11, DBCP, 1} and \eqref{Eqn: KKT simplified, 11, DBCP, 2} imply:
\begin{align*}
    \ell(2) = \lambda^E \leq \ell(0) + \frac{(1-\alpha)\tau}{v^E},
\end{align*}
a contradiction, since $\tau < v^E \big( \ell(2) - \ell(0) \big)$ by assumption. Thus, $y_1^E \ne 1$ and $y_2^E \ne 1$, so $y_1^E, y_2^E \in (0, 1)$. As a result:
\begin{align*}
    \ell(y_1^E) + \frac{(1-\alpha)\tau}{v^E} = \lambda^E = \ell(2 - y_1^E),
\end{align*}
i.e., $y^D(\alpha) = y_1^E = y^\dagger(\alpha)$, as desired.

\subsection{Proof of Theorem \ref{Thm: yc vs. yd, 11}}
\label{subsec: App A2, Proof of Thm yc vs. yd, 11}


As in the proof of Theorem \ref{Thm: yc, yd, 11}, for each $\alpha \in [0, 1]$, let $y^\dagger(\alpha)$ denote the unique solution to the fixed-point equation:
\begin{align} \nonumber
    \ell(y^\dagger(\alpha)) + \frac{(1-\alpha)\tau}{v^E} = \ell(2 - y^\dagger(\alpha)).
\end{align}
Since $y^D(\alpha) = 0$ for each $\alpha \in [0, 1 - v^E/v^I)$ and $y^D(\alpha) = y^\dagger(\alpha)$ for each $\alpha \in (1 - v^E/v^I, 1]$, it suffices to establish the following claim to establish Theorem \ref{Thm: yc vs. yd, 11}---If $\tau \leq 2v^E \ell'(1)$, then $y^\dagger(\alpha) \geq \alpha$ for each $\alpha \in [0, 1)$; if $\tau > 2v^E \ell'(1)$, then there exists a unique $\alpha_3 \in [0, 1]$ such that $y^\dagger(\alpha) > \alpha$ for each $\alpha \in [0, \alpha_3)$, and $y^\dagger(\alpha) < \alpha$ for each $\alpha \in (\alpha_3, 1)$.

First, suppose $\tau \leq 2v^E \ell'(1)$. Then, substituting $\alpha = 1$ into \eqref{Eqn: y dagger, first derivative}, we obtain:
\begin{align*}
    \frac{dy^\dagger}{d\alpha}(1) &= \frac{\tau}{v^E \big[\ell'(2 -y^\dagger(1)) + \ell'(y^\dagger(1)) \big]} = \frac{\tau}{2 v^E \ell'(1)} \leq 1.
\end{align*}
Since $y^\dagger$ is convex, $\frac{dy^\dagger}{d\alpha}(\alpha)$ increases in $\alpha$, and thus for any $\alpha \in [0, 1)$:
\begin{align*}
    y^\dagger(\alpha) &= y^\dagger(1) - \int_\alpha^1 \frac{dy^\dagger}{d\alpha}(\alpha) \ d\alpha \geq 1 - (1-\alpha) = \alpha,
\end{align*}
as desired. Next, suppose $\tau > 2v^E \ell'(1)$. Then $\frac{dy^\dagger}{d\alpha}(1) > 1$, and so there exists some $\delta > 0$ such that for each $\alpha \in [1-\delta, 1]$, we have $\frac{dy^\dagger}{d\alpha}(\alpha) > 1$. As a result:
\begin{align*}
    y^\dagger(1-\delta) = y^\dagger(1) - \int_{1-\delta}^1 \frac{dy^\dagger}{d\alpha}(\alpha) \ d\alpha < 1 - \big( 1 - (1 - \delta) \big) = 1-\delta.
\end{align*}
We now show that there exists a unique $\alpha_3 \in (0, 1-\delta)$ such that $y^\dagger(\alpha_3) = \alpha_3$. To show existence, let $f_2: [0, 1] \ra \R$ be given by $f_2(\alpha) := y^\dagger(\alpha) - \alpha$ for each $\alpha \in [0, 1]$. Then $f_2$ is continuous, $f_2(0) = y^\dagger(0) > 0$, and $f_2(1-\delta) = y^\dagger(1 - \delta) - (1 - \delta) < 0$. Thus, by the Intermediate Value Theorem, there exists some $\alpha_3 \in (0, 1-\delta)$ such that $f_3(\alpha_3) = 0$, i.e., such that $y^\dagger(\alpha_3) = \alpha_3$. To show uniqueness, suppose by contradiction that there exist $\alpha_3, \bar \alpha_3 \in (0, 1-\delta)$, with $\alpha_3 < \bar \alpha_3$ (without loss of generality), satisfying $y^\dagger(\alpha_3) = \alpha_3$ and $y^\dagger(\bar \alpha_3) = \bar \alpha_3$. Then, since $y^\dagger$ is strictly convex, we have:
\begin{align*}
    y^\dagger(\bar \alpha_3) &< \frac{1 - \bar \alpha_3}{1 - \alpha_3} y^\dagger(\alpha_3) + \frac{\bar \alpha_3 - \alpha_3}{1 - \alpha_3} y^\dagger(1).
\end{align*}
Substituting $y^\dagger(\bar \alpha_3) = \bar \alpha_3$, $y^\dagger(\alpha_3) = \alpha_3$, and rearranging terms, we find that $y^\dagger(1) > 1$, a contradiction. Thus, $\alpha_3$ is unique; moreover, $y^\dagger(\alpha) > \alpha$ for each $\alpha \in [0, \alpha_3)$, and $y^\dagger(\alpha) < \alpha$ for each $\alpha \in (\alpha_3, 1)$.


\end{document}